\documentclass[article]{mesa-NS}
\usepackage{times,mathptm}
\usepackage{amsmath,amsfonts,amssymb}
\usepackage{graphicx,graphics,epsfig}
\usepackage{fancyhdr,fancybox}
\usepackage{verbatim}
\usepackage{color}

\RequirePackage[T1]{fontenc}
\RequirePackage{epsf}
\RequirePackage{psfig}
\RequirePackage{mathptmx}      
\RequirePackage{flushend}
\RequirePackage[numbers]{natbib}
\RequirePackage[colorlinks,citecolor=blue,urlcolor=blue,linkcolor=blue]{hyperref}

\jvol{22}
\jnum{1}
\jyear{2015}

\numberwithin{equation}{section}
\numberwithin{theorem}{section}
\numberwithin{definition}{section}
\numberwithin{lemma}{section}
\numberwithin{corollary}{section}
\numberwithin{proposition}{section}
\numberwithin{example}{section}
\numberwithin{remark}{section}

\begin{document}
\label{pageinit}

\date{}

\title{Classifying orbits in a new dynamical model describing motion in a prolate or an oblate elliptical galaxy}

\author{Euaggelos E. Zotos $^{1^\star}$, Nicolaos D. Caranicolas $^1$, Efthimia G. Doni $^2$, }

\maketitle

\noindent $^1$ Department of Physics, Section of Astrophysics, Astronomy and Mechanics,\\
           Aristotle University of Thessaloniki, \\
           GR-541 24, Thessaloniki, Greece\\

\noindent $^2$ Department of Physics, Aristotle University of Thessaloniki,\\
           GR-541 24, Thessaloniki, Greece\\

\noindent $^\star$ \emph{Corresponding Author}. E-mail address: evzotos@physics.auth.gr

\markboth{}{Classifying orbits in a new dynamical model}

\footnotetext[2010]{\textit{{\bf Mathematics Subject Classification}}: Primary:  \\
{\bf Keywords}: . }

\begin{abstract}
The regular or chaotic character of orbits of stars moving in the meridional plane $(R,z)$ of an axially symmetric elliptical galaxy with a dense, massive spherical nucleus and a dark matter halo component is under investigation. In particular, we explore how the flattening of an elliptical galaxy influences the overall orbital structure of the system, by computing in each case the percentage of chaotic orbits, as well as the percentages of orbits composing the main regular families. In an attempt to discriminate safely and with certainty between regular and chaotic motion, we use the Smaller ALingment Index (SALI) method to extensive samples of orbits obtained by integrating numerically the basic equations of motion as well as the variational equations. In addition, a technique which is based mainly on the field of spectral dynamics that utilizes the Fourier transform of the time series of each coordinate is used for classifying the regular orbits into different families and also to recognize the secondary resonances that usually bifurcate from them. Three cases are considered in our work: (i) the case where the elliptical galaxy is prolate (ii) the case where a spherically symmetric elliptical galaxy is present and (iii) the case where the elliptical galaxy has an oblate shape. Comparison between the current results and early related work is also made.
\end{abstract}


\section{Introduction}
\label{intro}

Elliptical galaxies appear as bright systems of elliptical shape in the night sky. They contain small amounts of gas and dust and consist mainly of old stars. Observations show that the shape of elliptical galaxies can be prolate or oblate or even triaxial \cite{AR02,VR05}. It is known, that earlier astronomers believed that elliptical galaxies owe their flattening in their rotation which currently do not apply after the work of \cite{BC75} and \cite{I77}. Today, it is commonly believed, that the shape of elliptical galaxies is not due to rotation, but due to the anisotropic distribution of velocities of the stars \cite{B80a,B80b}. The formation and the evolution of elliptical galaxies can be described and interpreted by using suitable galactic models \cite{T77,WR78,M92,KB03,BBM06,dLS06,HHC09} in which the formation of the elliptical galaxies is explained using mergers \cite{MSC08,vW08,NJO09}.

Elliptical galaxies exhibit a variety of sizes. There are dwarfs, giants as well as normal elliptical galaxies. Dwarf elliptical galaxies can be found in clusters of galaxies such as the Fornax and the Virgo cluster but also exist in the Local Group. Two well known dwarf elliptical galaxies in the Local Group are the M32 and the NGC 205, that is the satellites of the Andromeda galaxy. Giant elliptical galaxies on the other hand, are considered as the most massive objects in the universe. Some astronomers believe, that giant elliptical galaxies were formed from the mergers of disk galaxies \cite[e.g.,][]{NJJ08}. One of the most famous elliptical galaxies is the M87 located near the centre of the Virgo cluster.

To help the reader, we quote some examples of academic papers over the last two decades related to the dynamics of elliptical galaxies. \cite{MQ98} explored the dynamical evolution of elliptical galaxies with central singularities, while \cite{FP99} studied the relationship between age and dynamics in elliptical galaxies. The internal dynamics, structure formation and evolution of dwarf elliptical galaxies was investigated by \cite{GGvdM03}. Dynamical models of elliptical galaxies were constructed and studied by \cite{J09}, while \cite{vv07} considered the mass-to-light ratio evolution of early-type galaxies using dynamical modelling of resolved internal kinematics. Furthermore, evolution of  elliptical galaxies induced by dynamical friction was examined by \cite{AB07}, who also studied the role of density concentration and pressure anisotropy on the galactic evolution.

Valuable information about the structure and the evolution of elliptical galaxies can be obtained from the study of the orbits of stars. In order to do that, it is necessary to build a dynamical model that describes the main orbital properties of the galaxy. One of the most classical models used for the description of axially symmetrical elliptical galaxies is the  logarithmic potential \cite[e.g.,][]{R80,BS82}
\begin{equation}
V_{\rm L} = \frac{\upsilon_0^2}{2} \ln\left(R^2 + \frac{z^2}{q^2} + c^2 \right),
\label{plog}
\end{equation}
where $(R,z)$ are the usual cylindrical coordinates, while $\upsilon_0$, $q$ and $c$ are suitable parameters. Another useful dynamical model used for the description of disk and elliptical galaxies is the Miyamoto-Nagai (1975) \cite{MN75} potential
\begin{equation}
V_{\rm MN} = \frac{-G M}{\sqrt{R^2 + \left(\alpha + \sqrt{h^2 + z^2}\right)^2}},
\label{pMN}
\end{equation}
where $M$ is the mass of the galaxy, $G$ is the gravitational constant, while $\alpha$ and $h$ are parameters controlling the shape of the galaxy. In fact, the potential of Eq. (\ref{pMN}) can describe an elliptical galaxy when $h \gg \alpha$, while when $\alpha \gg h$ describes a flat disk galaxy. The reader can find additional interesting information on models of galaxies in \cite{BT08}. Moreover, interesting information on the dynamics of elliptical galaxies can be found in the review paper of \cite{M99}.

Over the years, a huge load of research work has been devoted on determining the regular or chaotic nature of orbits in axisymmetric galaxies. However, the vast majority deals mostly with the discrimination between regular and chaotic motion, while only a tiny fraction of the existed literature proceeds further classifying orbits in different regular families. \cite{LS92}, in a thorough pioneer study, analyzed the orbital content in the coordinate planes of triaxial potentials but also in the meridional plane of axially symmetric model potentials, focusing on the regular families. Few years later, \cite{CA98} developed a method based on the analysis of the Fourier spectrum of the orbits which can distinguish not only between regular and chaotic orbits, but also between loop, box, and other resonant orbits either in two or three dimensional potentials. This spectral method was improved and applied in \cite{MCW05} in order to identify the different kinds of regular orbits in a self-consistent triaxial model. The same code was improved even further in \cite{ZC13}, when the influence of the central nucleus and of the isolated integrals of motion (angular momentum and energy) on the percentages of orbits in the meridional plane of an axisymmetric galactic model composed of a disk and a spherical nucleus were investigated. In two recent papers \cite{CZ13} and \cite{ZCar13}, analytical dynamical models describing the motion of stars both in disk and elliptical galaxies containing dark matter were used in order to investigate how the presence and the amount of dark matter influences the regular or chaotic nature of orbits as well as the behavior of the different families of resonant orbits.

Taking into account all the above, there is no doubt, that knowing the overall orbital structure in elliptical galaxies is an issue of paramount importance. On this basis, it seems of particular interest, to introduce a new analytical mass model in order to investigate the properties of motion in prolate as well as in oblate elliptical galaxies and then investigate how the flattening of the elliptical galaxy affects the regular or chaotic nature of orbits as well as the behavior of the different families of orbits. Here we must point out, that the present article belongs to a series of papers \cite{ZC13,CZ13,ZCar13,ZCar14} that have as main objective the orbit classification (not only regular versus chaotic but also separating regular orbits into different regular families) in different galactic gravitational potentials. Thus, we decided to follow a similar structure and the same numerical approach to all of them.

The structure of the present paper is as follows: In Section \ref{galmod} we present a detailed description of the properties of our composite gravitational galactic model. All the different computational methods used in order to determine the character of orbits are described in Section \ref{compmeth}. In the following Section, we explore how the particular shape of the elliptical galaxy (prolate, spherical or oblate), defined by the flattening parameter, influences the character of the orbits. Our article ends with Section \ref{disc}, where the discussion and the conclusions of this research are presented.

\section{Properties of the galactic model}
\label{galmod}

The main objective of this work is to reveal how the flattening of an elliptical galaxy influences the character of orbits moving in the meridional plane of an axially symmetric galaxy model with a central and spherical nucleus with an additional dark matter halo component. For convenience, we shall use the usual cylindrical coordinates $(R, \phi, z)$, where $z$ is the axis of symmetry.

The total gravitational potential $\Phi(R,z)$ consists of three mass model components: the main elliptical galaxy potential $\Phi_{\rm g}$, the central, spherical component $\Phi_{\rm n}$ and the dark matter halo component $\Phi_{\rm h}$. For the description of properties of a flattened elliptical galaxy we propose the following model
\begin{equation}
\Phi_{\rm g}(R,z) = \frac{-G M_{\rm g}}{\alpha + \sqrt{R^2 + b z^2}}.
\label{Vg}
\end{equation}
Here $G$ is the gravitational constant, $M_{\rm g}$ is the mass of the elliptical galaxy, $\alpha$ is a softening parameter, while $b$ is the flattening parameter controlling the shape of the elliptical galaxy. This model is somehow a combination of two well known galaxy models: (a) the Plummer model \cite{P11} and (b) the Hernquist's model \cite{H91}. We have chosen this dynamical model for three main reasons: (i) it is a simple mass model with low computational cost at the numerical calculations; (ii) the model potential of Eq. (\ref{Vg}) can describe a wide variety of shapes of elliptical galaxies by suitably choosing the parameter $b$. In particular, when $0.1 \leq b < 1$ the galaxy is prolate, when $b = 1$ is spherical, while when $1 < b \leq 1.9$ is oblate; (iii) it gives us the opportunity to compare the results of the present model with results obtained from other dynamical models, in an attempt to derive some useful conclusions concerning the motion in elliptical galaxies. A similar potential was used in a recent work \cite{ZCar14} in order to model the properties of a non-spherical nucleus.

For the description of the spherically symmetric nucleus we use a Plummer potential \cite[e.g.,][]{BT08}
\begin{equation}
\Phi_{\rm n}(R,z) = \frac{-G M_{\rm n}}{\sqrt{R^2 + z^2 + c_{\rm n}^2}},
\label{Vn}
\end{equation}
where $M_{\rm n}$ and $c_{\rm n}$ are the mass and the scale length of the nucleus, respectively. This potential has been used successfully in the past in order to model and therefore interpret the effects of the central mass component in a galaxy \cite[see, e.g.][]{HN90,HPN93,Z12a,ZC13}. At this point, we must make clear that Eq. (\ref{Vn}) is not intended to represent the potential of a black hole nor that of any other compact object, but just the potential of a dense and massive nucleus therefore, any relativistic effects are out of the scope of this work. Finally, the dark matter halo is modelled by a similar spherically symmetric mass model potential
\begin{equation}
\Phi_{\rm h}(R,z) = \frac{-G M_{\rm h}}{\sqrt{R^2 + z^2 + c_{\rm h}^2}},
\label{Vh}
\end{equation}
where $M_{\rm h}$ and $c_{\rm h}$ are the mass and the scale length of the halo, respectively. The spherical shape of the dark halo is simply an assumption, due to the fact that galactic halos may have a variety of shapes.

In our study, we use the well-known system of galactic units, where the unit of length is 1 kpc, the unit of mass is $2.325 \times 10^7 {\rm M}_\odot$ and the unit of time is $0.9778 \times 10^8$ yr. The velocity units is 10 km/s, the unit of angular momentum (per unit mass) is 10 km kpc s$^{-1}$, while $G$ is equal to unity. The energy unit (per unit mass) is 100 km$^2$s$^{-2}$. In these units, the values of the involved parameters are: $M_{\rm g} = 8000$ (corresponding to 1.86 $\times$ $10^{11}$ M$_{\odot}$), $\alpha = 10$, $M_n = 400$ (corresponding to 9.3 $\times$ $10^{9}$ M$_{\odot}$), $c_{\rm n} = 0.25$, $M_{\rm h} = 15000$ (corresponding to 3.48 $\times$ $10^{11}$ M$_{\odot}$) and $c_{\rm h} = 15$. These values were chosen having in mind a normal sized elliptical galaxy \cite[e.g.,][]{AS91}. The flattening parameter $b$, on the other hand, is treated as a parameter and its value varies in the interval $0.1 \leq b \leq 1.9$.

\begin{figure}[!tH]
\includegraphics[width=\hsize]{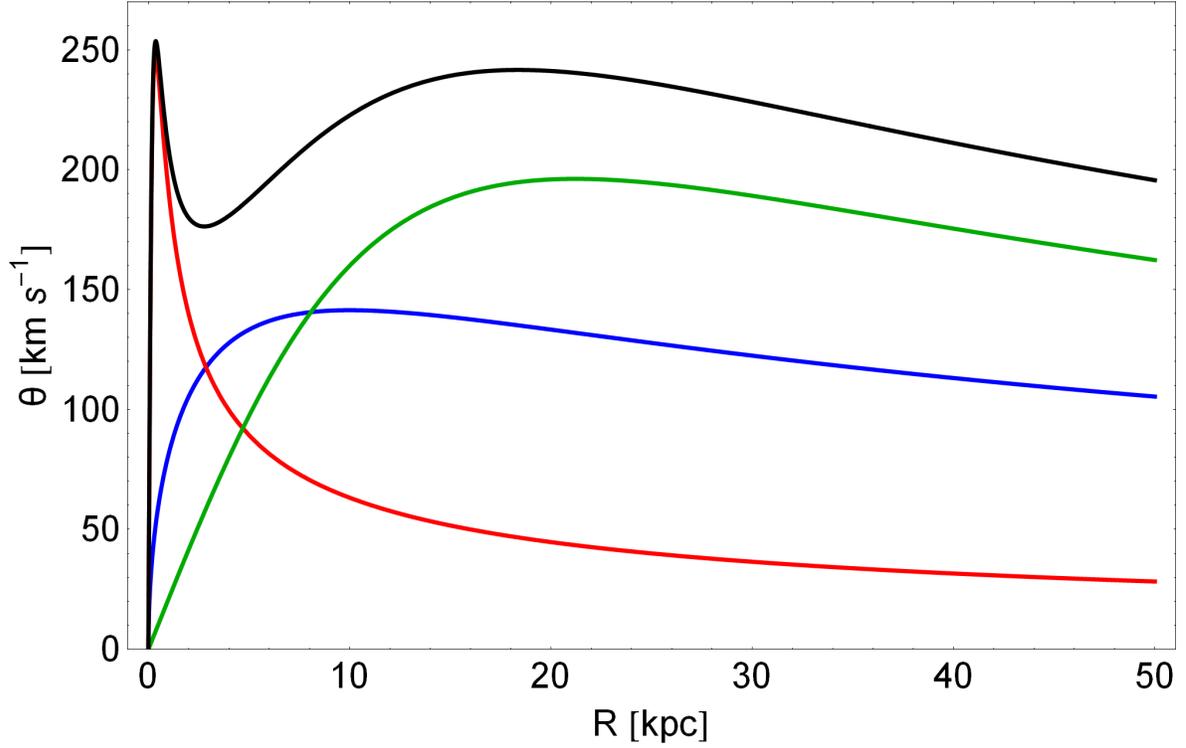}
\caption{A plot of the rotation curve in our galactic model. We can distinguish the total circular velocity (black) and also the contributions from the spherical nucleus (red), the elliptical galaxy's main body (blue) and that of the dark matter halo (green).}
\label{rotvel}
\end{figure}

\begin{figure*}[!tH]
\centering
\resizebox{0.9\hsize}{!}{\includegraphics{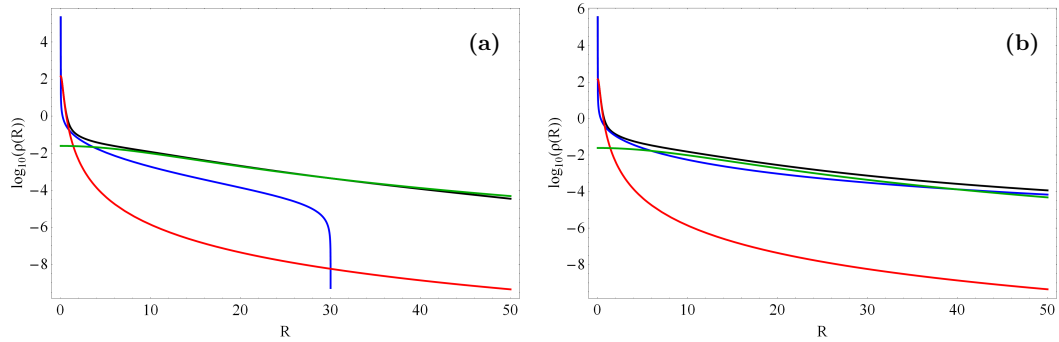}}
\caption{Evolution of the mass density $\rho(R)$ on the galactic plane $(z=0)$, as a function of the distance $R$ from the center when (a-left): $b = 0.5$ and (b-right): $b = 1.5$. The total mass density is shown in black color, while we can distinguish the three different contributions: the spherical nucleus (red), the elliptical galaxy's main body (blue) and the dark matter halo (green).}
\label{denR}
\end{figure*}

One of the most important parameters in axially symmetric galaxies is the circular velocity in the galactic plane $z=0$,
\begin{equation}
\theta(R) = \sqrt{R\left|\frac{\partial \Phi(R,z)}{\partial R}\right|_{z=0}}.
\label{cvel}
\end{equation}
Inserting the total potential $\Phi(R,z)$ in Eq. (\ref{cvel}) we obtain
\begin{equation}
\theta(R) = R \sqrt{G \left(\frac{M_{\rm g}}{R_{\rm g}} + \frac{M_{\rm n}}{R_{\rm n}} + \frac{M_{\rm h}}{R_{\rm h}}\right)},
\label{cvel1}
\end{equation}
where
\begin{eqnarray}
R_{\rm g} &=& R\left(\alpha + R\right)^2, \nonumber \\
R_{\rm n} &=& \left(R^2 + c_{\rm n}^2\right)^{3/2}, \nonumber \\
R_{\rm h} &=& \left(R^2 + c_{\rm h}^2\right)^{3/2}.
\label{Rs}
\end{eqnarray}
A plot of the total $\theta(R)$ for our galactic model is presented in Fig. \ref{rotvel}, as a black curve\footnote{On the galactic plane applies $z=0$ so, the rotation curve is the same regardless the prolate or oblate shape of the elliptical galaxy.}. Moreover, in the same plot, the red line shows the contribution from the spherical nucleus, the blue curve is the contribution from the elliptical galaxy's main body, while the green line corresponds to the contribution form the dark halo. It is seen, that each contribution prevails in different distances form the galactic center. In particular, at small distances when $R \leq 2.86$ kpc, the contribution from the spherical nucleus dominates, while at mediocre distances, $2.86 < R < 8.07$ kpc, the galaxy's main body contribution is the dominant factor. On the other hand, at relatively large galactocentric distances $R > 8.07$ kpc we see that the contribution from the dark halo prevails, thus forcing the circular velocity to possess high values with increasing distance from the center. We can also observe, the characteristic local minimum of the rotation curve due to the central and massive nucleus, which appears at small values of $R$, when fitting observed data to a galactic model \cite[e.g.,][]{GHBL10,IWTS13}.

\begin{figure}[!tH]
\includegraphics[width=\hsize]{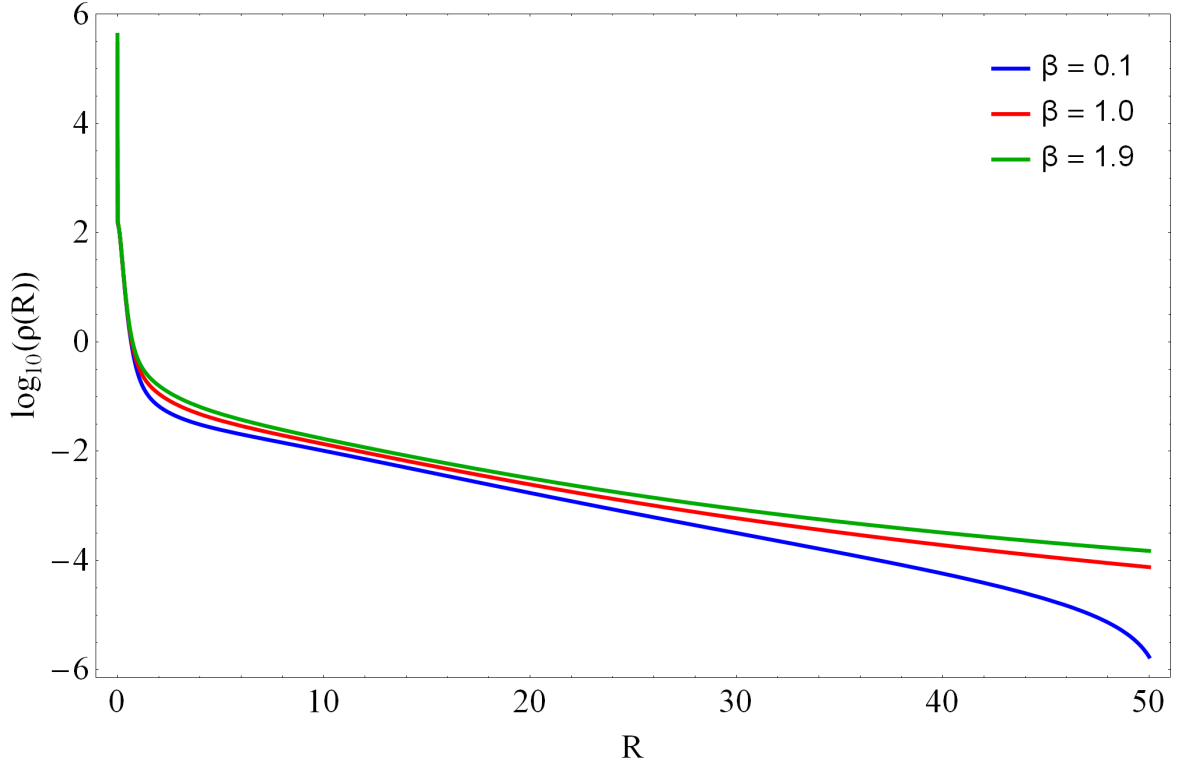}
\caption{Evolution of the total mass density $\rho(R)$ on the galactic plane $(z = 0)$, as a function of the distance $R$ from the center for three different values of the flattening parameter $b$. Details are given in the text.}
\label{denb}
\end{figure}

\begin{figure}[!tH]
\includegraphics[width=\hsize]{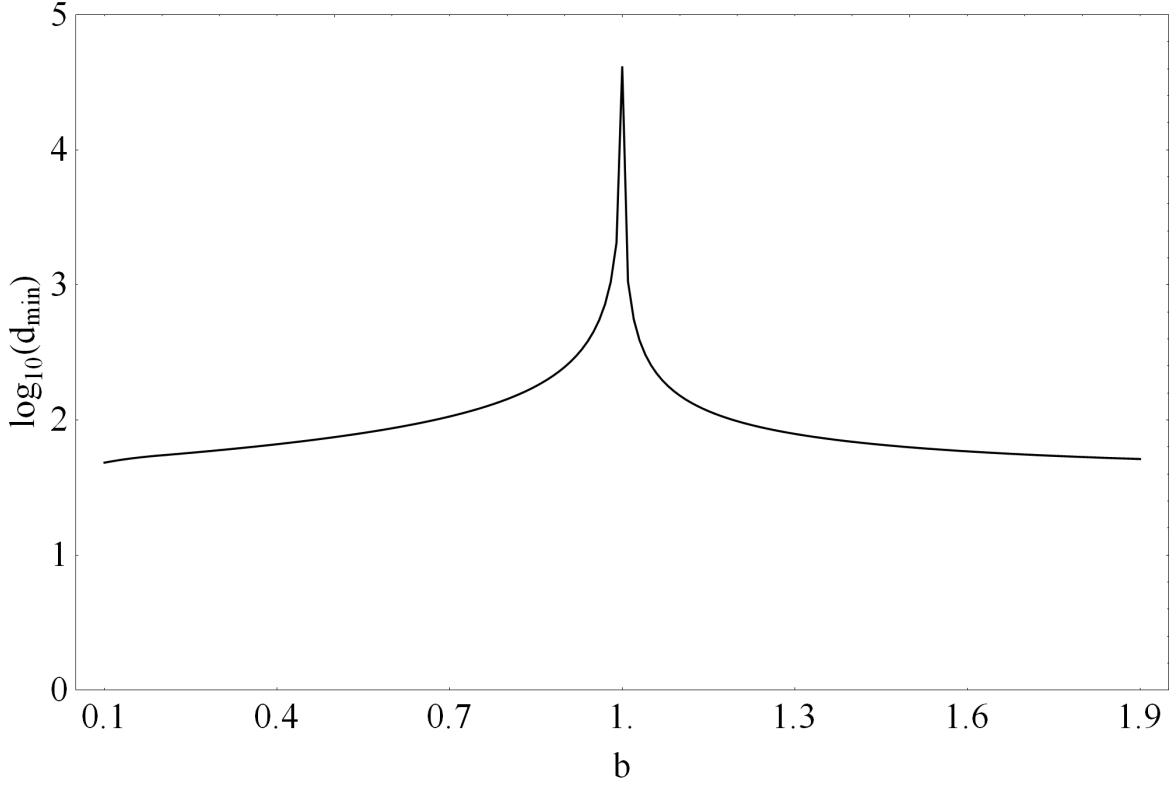}
\caption{The evolution of the minimum distance $d_{\rm min}$ where negative density appears for the first time, as a function of the flattening parameter $b$.}
\label{deneg}
\end{figure}

It is often very useful to compute the mass density $\rho(R,z)$ derived from the total potential $\Phi(R,z)$ using the Poisson's equation
\begin{eqnarray}
\rho(R,z) = \frac{1}{4 \pi G} \nabla^2 \Phi(R,z) = \nonumber \\
= \frac{1}{4 \pi G} \left(\frac{\partial^2}{\partial R^2} + \frac{1}{R}\frac{\partial}{\partial R}
+ \frac{1}{R^2} \frac{\partial^2}{\partial \phi^2} + \frac{\partial^2}{\partial z^2}\right) \Phi(R,z).
\label{dens}
\end{eqnarray}
Remember, that due to the axial symmetry the third term in Eq. (\ref{dens}) is zero. If we set $z = 0$ in Eq. (\ref{dens}) we obtain the mass density on the galactic plane which is
\begin{equation}
\rho(R) = \frac{1}{4\pi}\left(\rho_{\rm g} + \rho_{\rm n} + \rho_{\rm h}\right),
\label{densR}
\end{equation}
where
\begin{eqnarray}
\rho_{\rm g} &=& \frac{M_{\rm g}\left(\alpha\left(b + 1\right) + R\left(b - 1\right)\right)}{R\left(\alpha + R\right)^3}, \nonumber \\
\rho_{\rm n} &=& \frac{3M_{\rm n}c_{\rm n}^2}{\left(R^2 + c_{\rm n}^2\right)^{5/2}}, \nonumber \\
\rho_{\rm h} &=& \frac{3M_{\rm h}c_{\rm h}^2}{\left(R^2 + c_{\rm h}^2\right)^{5/2}}.
\label{densR2}
\end{eqnarray}
In the following Fig. \ref{denR}(a-b) the evolution of the total mass density $\rho(R,z=0)$ on the galactic plane, as a function of the of the radius $R$ from the galactic center is shown as the black curve. Fig. \ref{denR}a corresponds to a prolate elliptical galaxy with $b = 0.5$, while Fig. \ref{denR}b to the case of an oblate elliptical galaxy with $b = 1.5$. In both plots, the red line shows the contribution from the spherical nucleus, the blue curve is the contribution from the elliptical galaxy's main body, while the green line corresponds to the contribution form the dark matter halo. It is evident, that the density of the nucleus decreases rapidly obtaining very low values, while the density of the dark halo continues to hold significantly larger values coinciding with the total density. The density of the elliptical galaxy's main body on the other hand, behaves completely differently in both cases. In particular, in the case of the prolate elliptical galaxy the density $\rho_{\rm g}$ becomes effectively zero when $R > 30$ kpc, while in the case of the oblate elliptical it continues to exist even at large galactocentric distances.

In Fig. \ref{denb} we present the evolution of the total $\rho(R)$ as a function of the of the radius $R$ from the galactic center, for three values of the flattening parameter $b$. In the same diagram, the blue line correspond to $b = 0.1$, the red curve to $b = 1$, while the green line corresponds to $b = 1.9$. We observe, that there are no significant deviations at the evolution of the density and only at large galactocentric distances ($R > 30$ kpc) we may say that the value of the flattening parameter $b$ influences the mass density. According to our numerical calculations, at large galactocentric distances the total mass density varies like $R^{-k}$, where $3.95 \leq k \leq 5.21$ when $0.1 \leq b \leq 0.9$, $k = 3.87$ when $b = 1$ and $3.79 \leq k \leq 3.42$ when $1.1 \leq b \leq 1.9$.

At this point, we must clarify, that the mass density in our elliptical galaxy model obtains negative values when the distance from the centre of the galaxy described by the model exceeds a minimum distance $d_{min} = \sqrt{R^2 + b z^2}$, which strongly depends on the flattening parameter $b$. Fig. \ref{deneg} shows a plot of the minimum distance $d_{\rm min}$ where the first indication of $\rho < 0$ takes place versus $b$ for our elliptical galaxy models. We see, that even at the extreme values of the flattening parameter, that is when $b = 0.1$ and $b = 1.9$, the areas of negative density occurs only when $d_{\rm min} > 50$ kpc, that is almost at the theoretical boundaries of a real galaxy. It must be pointed out, that our gravitational potential is truncated at $R_{\rm max} = 50$ kpc in order to avoid the existence of negative values of density.

If we take into account that our total gravitational potential $\Phi(R,z)$ is axially symmetric, then the $z$-component of the angular momentum $L_z$ is conserved. With this restriction, orbits can be described by means of the effective potential
\begin{equation}
\Phi_{\rm eff}(R,z) = \Phi(R,z) + \frac{L_z^2}{2R^2}.
\label{veff}
\end{equation}
In Eq. (\ref{veff}) the $L_z^2/(2R^2)$ term acts as a centrifugal barrier, allowing only orbits with sufficiently small $L_z$ to pass near the axis of symmetry. Therefore, the three-dimensional (3D) motion is effectively reduced to a two-dimensional (2D) motion in the meridional plane $(R,z)$, which rotates non-uniformly around the axis of symmetry according to
\begin{equation}
\dot{\phi} = \frac{L_z}{R^2},
\label{rotv}
\end{equation}
where of course the dot indicates derivative with respect to time.

On the meridional plane the equations of motion take the form
\begin{equation}
\ddot{R} = - \frac{\partial \Phi_{\rm eff}}{\partial R}, \ \ \
\ddot{z} = - \frac{\partial \Phi_{\rm eff}}{\partial z},
\label{eqmot}
\end{equation}
while the equations describing the evolution of a deviation vector $\delta {\bf{w}} = (\delta R, \delta z, \delta \dot{R}, \delta \dot{z})$ which joins the corresponding phase space points of two initially nearby orbits, needed for the calculation of standard chaos indicators (the SALI in our case) are given by the following variational equations
\begin{eqnarray}
\dot{(\delta R)} &=& \delta \dot{R}, \ \ \
\dot{(\delta z)} = \delta \dot{z}, \nonumber \\
(\dot{\delta \dot{R}}) &=&
- \frac{\partial^2 \Phi_{\rm eff}}{\partial R^2} \delta R
- \frac{\partial^2 \Phi_{\rm eff}}{\partial R \partial z}\delta z,
\nonumber \\
(\dot{\delta \dot{z}}) &=&
- \frac{\partial^2 \Phi_{\rm eff}}{\partial z \partial R} \delta R
- \frac{\partial^2 \Phi_{\rm eff}}{\partial z^2}\delta z.
\label{vareq}
\end{eqnarray}

Consequently, the corresponding Hamiltonian to the effective potential given in Eq. (\ref{veff}) reads
\begin{equation}
H = \frac{1}{2} \left(\dot{R}^2 + \dot{z}^2 \right) + \Phi_{\rm eff}(R,z) = E,
\label{ham}
\end{equation}
where $\dot{R}$ and $\dot{z}$ are momenta per unit mass, conjugate to $R$ and $z$ respectively, while $E$ is the numerical value of the Hamiltonian, which is conserved. Thus, all orbits are restricted to the area in the meridional plane satisfying $E \geq V_{\rm eff}$.

\section{Description of the computational methods}
\label{compmeth}

Knowing whether an orbit is regular or chaotic is an issue of paramount importance. Over the years, several chaos indicators have been developed in order to determine the nature of orbits. In our investigation, we chose to use the Smaller ALingment index (SALI) method. The SALI has been proved a very fast, reliable and effective tool, which is defined as
\begin{equation}
\rm SALI(t) = min(d_-, d_+),
\label{sali}
\end{equation}
where $d_- \equiv \| {\bf{w_1}}(t) - {\bf{w_2}}(t) \|$ and $d_+ \equiv \| {\bf{w_1}}(t) + {\bf{w_2}}(t) \|$ are the alignments indices, while ${\bf{w_1}}(t)$ and ${\bf{w_2}}(t)$, are two deviations vectors which initially point in two random directions. For distinguishing between ordered and chaotic motion, all we have to do is to compute the SALI for a relatively short time interval of numerical integration $t_{max}$. In particular, we track simultaneously the time-evolution of the main orbit itself as well as the two deviation vector ${\bf{w_1}}(t)$ and ${\bf{w_2}}(t)$ in order to compute the SALI. The variational equations (\ref{vareq}), as usual, are used for the evolution and computation of the deviation vectors.

\begin{figure}[!tH]
\includegraphics[width=\hsize]{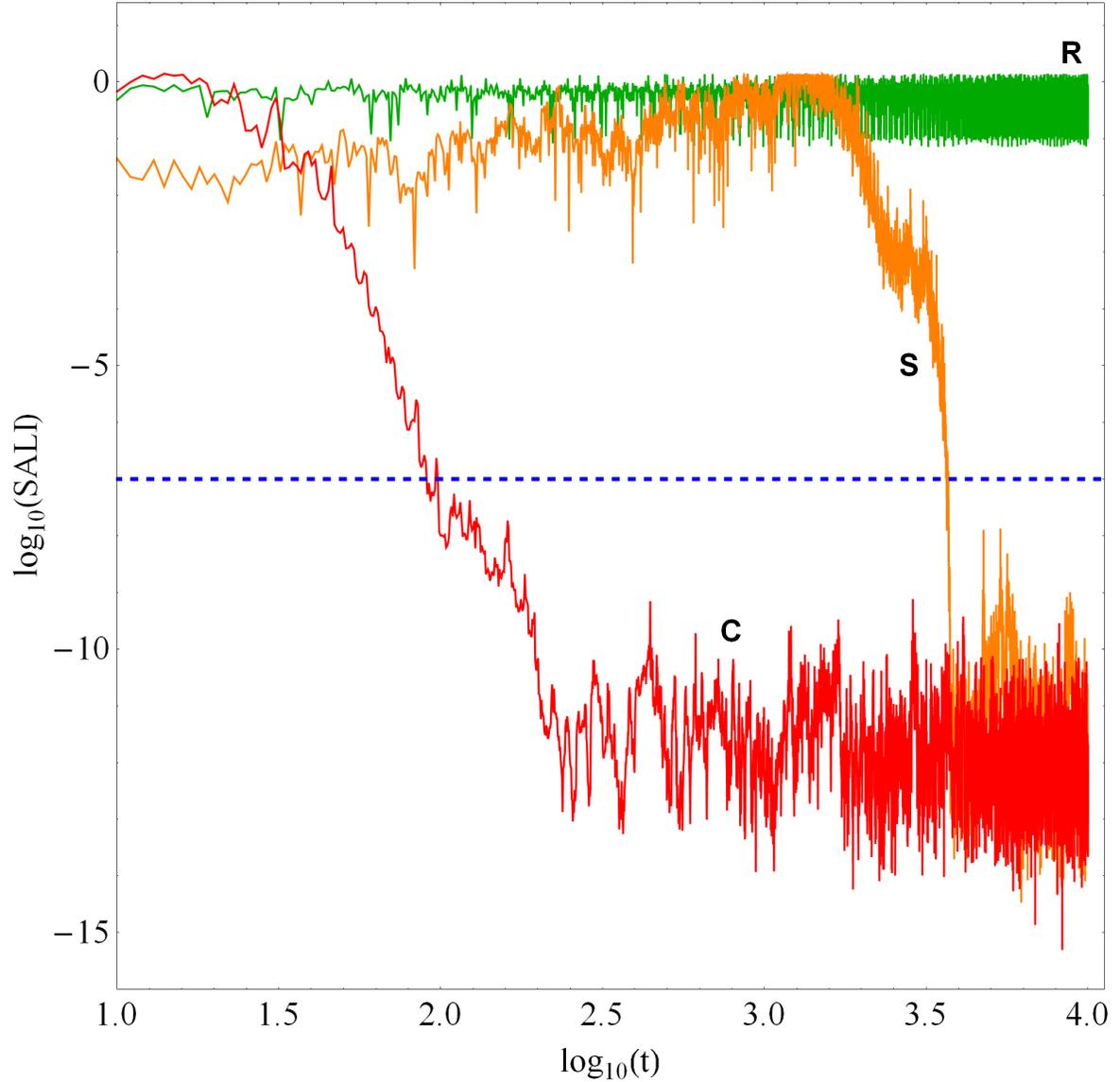}
\caption{Time-evolution of the SALI of a regular orbit (green color - R), a sticky orbit (orange color - S) and a chaotic orbit (red color - C) in our model for a time period of $10^4$ time units. The horizontal, blue, dashed line corresponds to the threshold value $10^{-7}$ which separates regular from chaotic motion. The chaotic orbit needs only about 90 time units in order to cross the threshold, while on the other hand, the sticky orbit requires a considerable longer integration time of about 3600 time units so as to reveal its true chaotic nature.}
\label{SALIevol}
\end{figure}

\begin{figure}[!tH]
\includegraphics[width=\hsize]{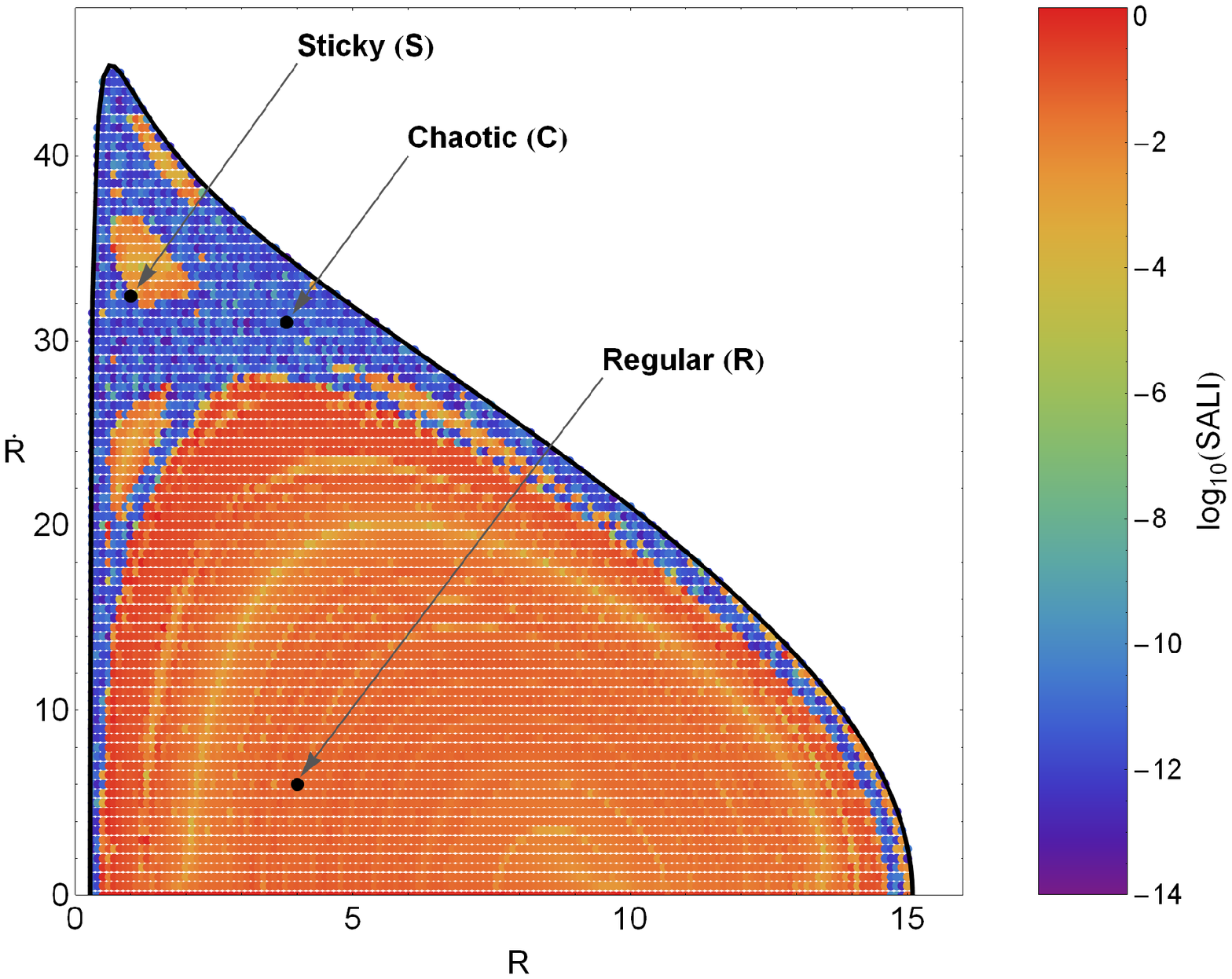}
\caption{Regions of different values of the SALI on the $(R,\dot{R})$ phase plane when the flattening parameter is $b = 1.5$. Light reddish colors correspond to ordered motion, dark blue/purplr colors indicate chaotic motion, while all intermediate colors suggest sticky orbits. The three black dots point the initial conditions of the orbits of Fig. \ref{SALIevol}. For more details regarding the grid structure see subsection \ref{obl}.}
\label{SALIgrid}
\end{figure}

The time-evolution of SALI strongly depends on the nature of the computed orbit since when the orbit is regular the SALI exhibits small fluctuations around non zero values, while on the other hand, in the case of chaotic orbits the SALI after a small transient period it tends exponentially to zero approaching the limit of the accuracy of the computer $(10^{-16})$. Therefore, the particular time-evolution of the SALI allow us to distinguish fast and safely between regular and chaotic motion. The time-evolution of a regular (R) and a chaotic (C) orbit for a time period of $10^4$ time units is presented in Fig. \ref{SALIevol}. We observe, that both regular and chaotic orbits exhibit the expected behavior. Nevertheless, we have to define a specific numerical threshold value for determining the transition from regularity to chaos. After conducting extensive numerical experiments, integrating many sets of orbits, we conclude that a safe threshold value for the SALI taking into account the total integration time of $10^4$ time units is the value $10^{-7}$. The horizontal, blue, dashed line in Fig. \ref{SALIevol} corresponds to that threshold value which separates regular from chaotic motion. In order to decide whether an orbit is regular or chaotic, one may use the usual method according to which we check after a certain and predefined time interval of numerical integration, if the value of SALI has become less than the established threshold value. Therefore, if SALI $\leq 10^{-7}$ the orbit is chaotic, while if SALI $ > 10^{-7}$ the orbit is regular. In Fig. \ref{SALIgrid} we present a dense grid of initial conditions $(R_0,\dot{R_0})$, where the values of the SALI are plotted using different colors. We clearly observe several regions of regularity indicated by light reddish colors as well as a unified chaotic domain (blue/purple dots). All intermediate colors correspond to sticky orbits. The initial conditions of the three orbits (regular, sticky and chaotic) whose SALI time-evolution was given in Fig. \ref{SALIevol}, are pinpointed with black dots in Fig. \ref{SALIgrid}. Therefore, the distinction between regular and chaotic motion is clear and beyond any doubt when using the SALI method.

For determining the chaoticity of our models, we chose, for each set of values of the flattening parameter, a dense grid of initial conditions in the $(R,\dot{R})$ phase plane, regularly distributed in the area allowed by the value of the energy $E$. The points of the grid were separated 0.1 units in $R$ and 0.5 units in $\dot{R}$ direction. For each initial condition, we integrated the equations of motion (\ref{eqmot}) as well as the variational equations (\ref{vareq}) with a double precision Bulirsch-Stoer algorithm \cite[e.g.,][]{PTVF92}. In all cases, the energy integral (Eq. (\ref{ham})) was conserved better than one part in $10^{-10}$, although for most orbits, it was better than one part in $10^{-11}$.

In our study, each orbit was integrated numerically for a time interval of $10^4$ time units (10 billion yr), which corresponds to a time span of the order of hundreds of orbital periods but of the order of one Hubble time. The particular choice of the total integration time is an element of great importance, especially in the case of the so called ``sticky orbits" (i.e., chaotic orbits that behave as regular ones during long periods of time). A sticky orbit could be easily misclassified as regular by any chaos indicator\footnote{Generally, dynamical methods are broadly split into two types: (i) those based on the evolution of sets of deviation vectors in order to characterize an orbit and (ii) those based on the frequencies of the orbits which extract information about the nature of motion only through the basic orbital elements without the use of deviation vectors.}, if the total integration interval is too small, so that the orbit do not have enough time in order to reveal its true chaotic character. Thus, all the sets of orbits of a given grid were integrated, as we already said, for $10^4$ time units, thus avoiding sticky orbits with a stickiness at least of the order of a Hubble time. All the sticky orbits which do not show any signs of chaoticity for $10^4$ time units are counted as regular ones, since that vast sticky periods are completely out of scope of our research. A characteristic example of a sticky orbit (S) in our galactic model can be seen in Fig. \ref{SALIevol}, where we observe that the chaotic character of the particular sticky orbit is revealed only after a considerable long integration time of about 3600 time units.

A first step towards the understanding of the overall orbital behavior of our galactic system is knowing the regular or chaotic nature of orbits. In addition, of particular interest is the distribution of regular orbits into different families. Therefore, once the orbits have been characterized as regular or chaotic, we then further classified the regular orbits into different families, by using the frequency analysis of \cite{CA98}. Initially, \cite{BS82,BS84} proposed a technique, dubbed spectral dynamics, for this particular purpose. Later on, this method has been extended and improved by \cite{SN96} and \cite{CA98}. In a recent work, \cite{ZC13} the algorithm was refined even further so it can be used to classify orbits in the meridional plane. In general terms, this method calculates the Fourier transform of the coordinates of an orbit, identifies its peaks, extracts the corresponding frequencies and search for the fundamental frequencies and their possible resonances. Thus, we can easily identify the various families of regular orbits and also recognize the secondary resonances that bifurcate from them. The very same algorithm was used in all the papers of this series \cite{ZC13,CZ13,ZCar13,ZCar14}.

At the end of this Section, we would like to make a short note regarding the nomenclature of the orbits. All the orbits in an axially symmetric potential are in fact three-dimensional (3D) loop orbits, i.e., orbits that rotate around the axis of symmetry always in the same direction. However, in dealing with the meridional plane the rotational motion is lost, so the path that an orbit follows onto this plane can take any shape, depending on the nature of the orbit. Throughout this article, we will call an orbit according to its behaviour in the meridional plane. Therefore, if for example an orbit is a rosette lying in the equatorial plane of the axisymmetric potential, it will be a linear orbit in the meridional plane, etc.

\begin{figure*}[!tH]
\centering
\resizebox{0.8\hsize}{!}{\includegraphics{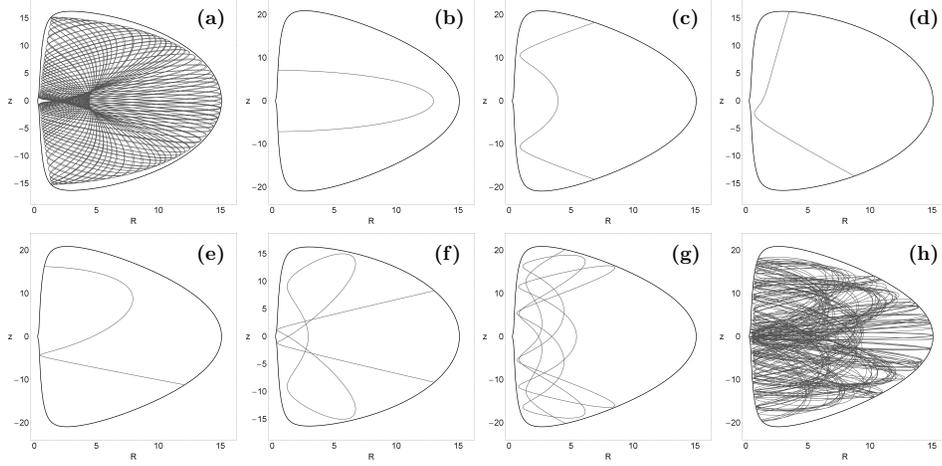}}
\caption{Orbit collection of the eight basic types in the prolate elliptical galaxy model: (a) box orbit; (b) 2:1 banana-type orbit (type a); (c) 2:1 resonant type b orbit; (d) 2:1 resonant type c orbit; (e) 3:1 boxlet orbit; (f) 6:3 boxlet orbit; (g) 10:5 boxlet orbit; (h) chaotic orbit.}
\label{orbsP}
\end{figure*}

\begin{figure*}[!tH]
\centering
\resizebox{0.8\hsize}{!}{\includegraphics{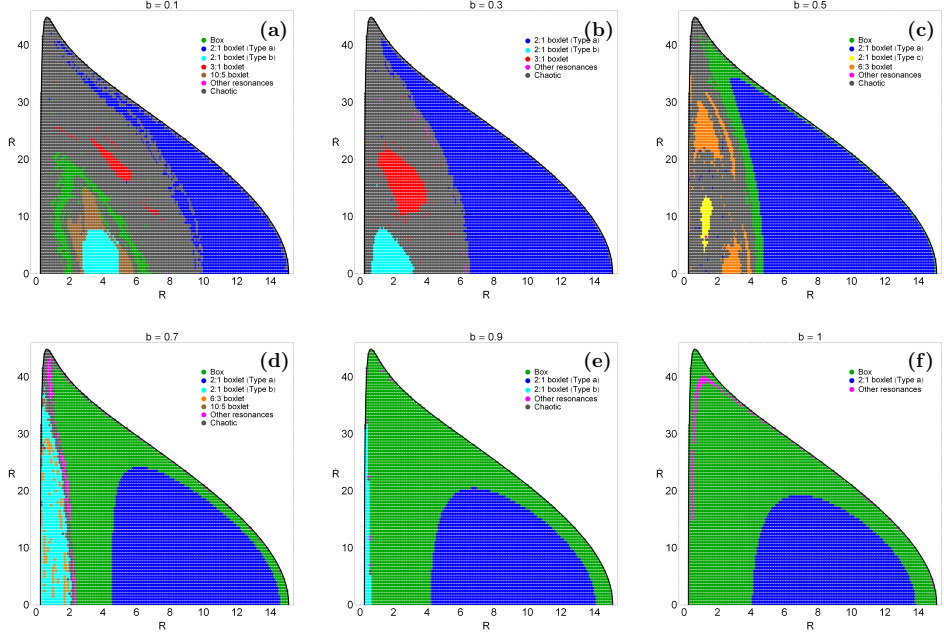}}
\caption{Orbital structure of the $(R,\dot{R})$ phase plane of the prolate elliptical galaxy model for different values of the flattening parameter $b$.}
\label{gridsP}
\end{figure*}

\section{Orbit classification \& Numerical results}
\label{orbclas}

This Section contains all the numerical results of our research. We numerically integrate several sets of orbits in order to distinguish between regular and chaotic motion. We use the initial conditions of orbits mentioned in Section \ref{compmeth} in order to construct the respective grids, taking always values inside the Zero Velocity Curve (ZVC) defined by
\begin{equation}
\frac{1}{2} \dot{R}^2 + \Phi_{\rm eff}(R,0) = E.
\label{zvc}
\end{equation}
In all cases, the value of the angular momentum of the orbits was set to $L_z = 15$ and kept constant. We chose for both prolate and oblate elliptical galaxy models the energy level -1050 which gives $R_{\rm max} \simeq 15$ kpc, where $R_{\rm max}$ is the maximum possible value of $R$ on the $(R,\dot{R})$ phase plane. Once the values of the parameters are chosen, we compute a set of initial conditions as described in Section \ref{compmeth} and integrate the corresponding orbits calculating the value of SALI and then classifying the regular orbits into different families. Each grid contains a total of 7702 initial conditions $(R_0,\dot{R_0})$ of orbits.

\subsection{Prolate elliptical galaxy model}
\label{prl}

\begin{table}
\begin{center}
   \caption{Type and initial conditions of the prolate elliptical galaxy model orbits shown in Figs. \ref{orbsP}(a-h). In all cases, $z_0 = 0$, $\dot{z_0}$ is found from the energy integral, Eq. (\ref{ham}), while $T_{\rm per}$ is the period of the resonant parent periodic orbits.}
   \label{table1}
   \setlength{\tabcolsep}{2.5pt}
   \begin{tabular}{@{}lcccc}
      \hline
      Figure & Type & $R_0$ & $\dot{R_0}$ & $T_{\rm per}$  \\
      \hline
      \ref{orbsP}a &  box        &  4.42000000 &  0.00000000 &           - \\
      \ref{orbsP}b &  2:1 type a & 13.00553103 &  0.00000000 &  3.06433481 \\
      \ref{orbsP}c &  2:1 type b &  3.95110793 &  0.00000000 &  4.45799389 \\
      \ref{orbsP}d &  2:1 type c &  1.30933861 & 10.48749236 &  3.61818580 \\
      \ref{orbsP}e &  3:1 boxlet &  4.68013558 & 19.38120702 &  4.12222325 \\
      \ref{orbsP}f &  6:3 boxlet &  2.89669399 &  0.00000000 & 10.72409920 \\
      \ref{orbsP}g & 10:5 boxlet &  5.43652122 &  0.00000000 & 22.25233730 \\
      \ref{orbsP}h & chaotic    &  0.24800000 &  0.00000000 &           - \\
      \hline
   \end{tabular}
\end{center}
\end{table}

The numerical calculations reveal, that when the elliptical galaxy has a prolate shape the orbital structure of the dynamical system is indeed very interesting. This is true because the 2:1 resonance appears with three different forms, while the 3:1 bifurcation is also present. Moreover, subfamilies of the main 2:1 family such as the 6:3 and 10:5 are also exist. An orbit with an improper ratio of frequencies (i.e., a ratio which is a reducible fraction) is a member of a subfamily of the orbits with a ratio of frequencies which is the irreducible corresponding fraction. For instance, always a 6:3 orbit torus surrounds a 2:1 torus; a 6:3 entire subfamily (as opposed to a single, invisible 6:3 orbit) appears when the parent 6:3 is stable and so it spawns its own subfamily. They have a parent closed 6:3 orbit the torus of which surrounds the 2:1 parent torus. Thus, in one sense the 6:3 orbits are the same as 2:1 ones (they are members of the same family), but in another sense they aren't the same (6:3 is a separated subfamily). In our research, we consider both 6:3 and 10:5 orbits to form separate families of orbits. It should be noticed, that basic resonant families, such as the 1:1, 3:2 and 4:3 are completely absent from the prolate elliptical galaxy models. Here we should like to clarify, that every resonance $n:m$ is expressed in such a way that $m$ is equal to the total number of islands of invariant curves produced in the $(R,\dot{R})$ phase plane by the corresponding orbit. In Fig. \ref{orbsP}(a-h) we present an example of each of the seven basic types of regular orbits, plus an example of a chaotic one. In orbits shown in Figs. \ref{orbsP}(a,d,f) we set $b = 0.5$, while in all other orbits the value of the flattening parameter is $b = 0.1$. The orbits shown in Figs. \ref{orbsP}a and \ref{orbsP}h were computed until $t = 100$ time units, while all the parent periodic orbits were computed until one period has completed. The black thick curve circumscribing each orbit is the limiting curve in the meridional plane $(R,z)$  defined as $\Phi_{\rm eff}(R,z) = E$. Table \ref{table1} gives the type and the initial conditions for all the depicted orbits. In the resonant cases, the initial conditions and the period $T_{\rm per}$ correspond to the parent periodic orbits.

\begin{figure}[!tH]
\includegraphics[width=\hsize]{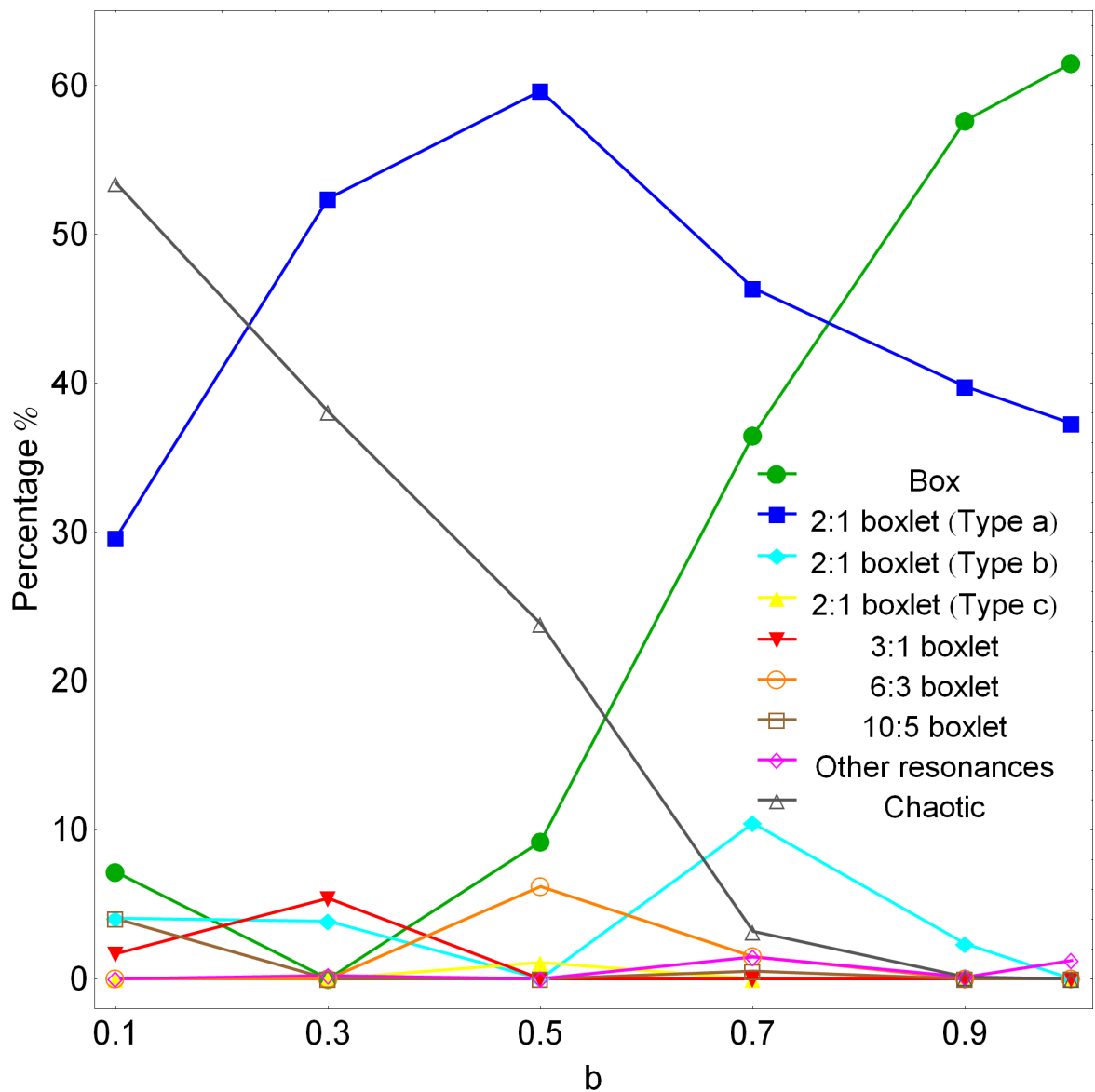}
\caption{Evolution of the percentages of the different kinds of orbits in our prolate elliptical galaxy model, when varying the flattening parameter $b$.}
\label{percsP}
\end{figure}

\begin{figure*}[!tH]
\centering
\resizebox{0.5\hsize}{!}{\includegraphics{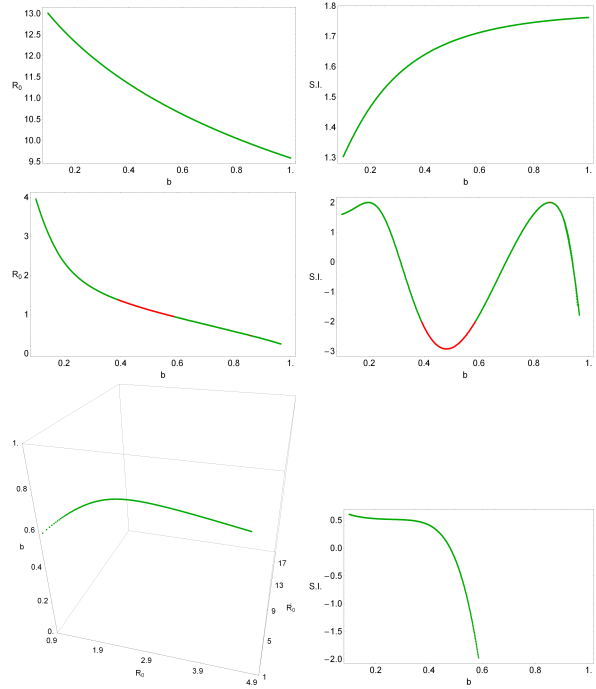}}
\caption{(left pattern): Evolution of the starting position $(R_0,\dot{R_0})$ of the 2:1 resonant periodic orbits as a function of the flattening parameter $b$. (right pattern): Evolution of the corresponding stability index (S.I.) of the families of periodic orbits as a function of the flattening parameter $b$. From top to bottom: 2:1 type a (banana-type) resonant family, 2:1 type b resonant family and 2:1 type c resonant family. Green dots indicate stable periodic orbits, while red dots correspond to unstable periodic orbits.}
\label{fpo21P}
\end{figure*}

\begin{figure*}[!tH]
\centering
\resizebox{0.5\hsize}{!}{\includegraphics{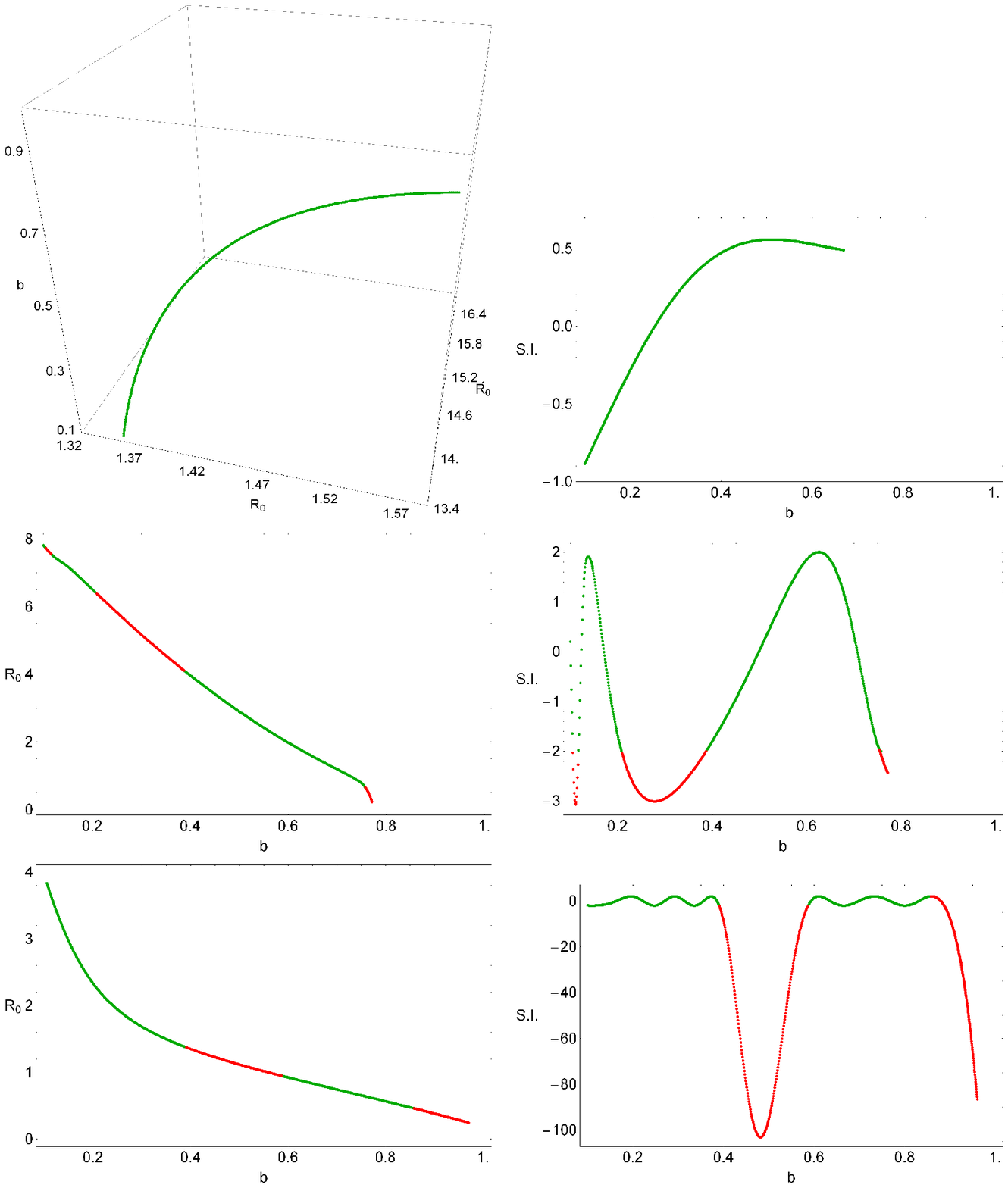}}
\caption{Similar to Fig. \ref{fpo21P}. From top to bottom: 3:1 resonant family, 6:3 resonant family and 10:5 resonant family. Green dots indicate stable periodic orbits, while red dots correspond to unstable periodic orbits.}
\label{fpoP}
\end{figure*}

To explore the influence of the flattening parameter $b$ of the elliptical galaxy on the orbital structure of the galactic system, we let it vary while fixing the numerical values of all the other parameters in our model. As already said, we fixed the values of all the other parameters and integrate orbits in the meridional plane for the set $b = \{0.1,0.3,0.5, ..., 1\}$. In all cases, the energy was set to $-1050$. Once the values of the parameters were chosen, we computed a set of initial conditions as described in Section \ref{compmeth} and integrated the corresponding orbits computing the SALI of the orbits and then classifying the regular orbits into different families.

As seen in Figs. \ref{orbsP}(b-d), three different types of 2:1 resonant orbits are present. However, all three types produce the same number of basic frequencies therefore, the orbit classification program is unable to distinguish between them. Nevertheless, we can exploit the shape of the orbits in order to split them in different categories. In particular, we observe that both orbits shown in Figs. \ref{orbsP}b and \ref{orbsP}c are symmetrical to the $R$ axis $(z = 0)$, while the orbit depicted in Fig. \ref{orbsP}d is not. Thus, we define the periodic orbit of Fig. \ref{orbsP}d as the mother orbit of the 2:1 resonant type c orbits. Similarly, the periodic orbits in Figs. \ref{orbsP}b and \ref{orbsP}c are the mother orbits of the 2:1 resonant type a and type b respectively. Both 2:1 type a and type b orbits are symmetrical to the $R$ axis so we have to find another separation criterion. Type a orbits are in fact the usual banana-type orbits which exhibit their maximum value of the $z$ coordinate $(z_{\rm max})$ at very low values of $R$, while on the contrary, type b orbits (which look like the Greek capital letter $\Sigma$), move at high departures from the galactic plane and reach the maximum value of $z$ at relatively large distances from the center. After extensive numerical tests, we concluded that a safe threshold value of the distance $R$ in which $z_{\rm max}$ is exhibited is $R_{\rm s} = 1.37$ kpc. Therefore, if $z_{\rm max}$ is obtained at $R \leq R_{\rm s}$ the 2:1 orbit is of type a, while if not it belongs to the type b category.

In Figs. \ref{gridsP}(a-f) we present six grids of initial conditions $(R_0,\dot{R_0})$ of orbits that we have classified for different values of the flattening parameter $b$. Here, we can identify all the different regular families by the corresponding sets of islands which are formed in the phase plane. In particular, we see the seven main families already mentioned: (i) 2:1 banana-type orbits surrounding the central main periodic point; (ii) 2:1 type b orbits located near the center; (iii) 2:1 type c resonant orbits forming a stability island for low values of $R$; (iv) 3:1 resonant orbits located near the center of the phase plane; (v) 6:3 resonant orbits producing a set of three islands near the 2:1 type b or c orbits; (vi) 10:5 resonant orbits corresponding to a chain of five small islands in the vicinity of the 2:1 type b orbits and (vii) box orbits situated mainly outside of the 2:1 type a resonant orbits. It is seen, that apart from the regions of regular motion, we observe the presence of a unified chaotic sea which embrace all the islands of stability. The outermost black thick curve is the ZVC defined by Eq. (\ref{zvc}).

When the elliptical galaxy has an extreme prolate shape, that is when $b = 0.1$, we see in Fig. \ref{gridsP}a that the majority of the phase plane is covered by initial conditions corresponding to chaotic orbits, while two different types of 2:1 resonant orbits coexist. A strong presence of the bifurcated 10:5 family is also observed around the 2:1 type b orbits. The structure of the phase plane changes when $b = 0.3$, where there is a complete absence of box and secondary orbits, while only basic resonances, such as the 2:1 types a and b and the 3:1 families exist. A much more radical mutation of the phase plane is observed in Fig. \ref{gridsP}c where the flattening parameter is $b = 0.5$. Here, the 2:1 type b and the 3:1 orbits disappear completely and the 2:1 type c and the 6:3 resonances emerge. We could also notice, the presence of box orbits creating a boundary between 2:1 type a and chaotic orbits. As the value of the flattening parameter increases $(b > 0.5)$ and the elliptical galaxy becomes less prolate tending to spherical symmetric, we see that the extent of the regions corresponding to chaotic and 2:1 (type a and b) orbits is reduced drastically and box orbits seem to take over the majority of the phase plane. A mixture of 2:1 type b, 6:3 and 10:5 orbits is present near the center when $b = 0.7$, while the 5:3 resonance (indicated as ``other resonances") produces a set of three small islands at the outer parts of the box region. Fig. \ref{gridsP}f shows the case of the spherically symmetric galaxy $(b = 1)$. As expected, there is no evidence of chaos and therefore, the 2:1 type a and box orbits share almost the entire phase plane. We say almost, because the 8:5 resonance makes its presence felt inside the vast region of box orbits.

Fig. \ref{percsP} shows the resulting percentages of the chaotic orbits and that of the main families of regular orbits as $b$ varies. It can be seen, that there is a strong correlation between the percentages of most types of orbits and the value of flattening parameter. When the elliptical galaxy is highly prolate $(b = 0.1)$ the orbital structure of the system exhibits great amount of chaos; more than 50\% of the phase plane is covered by chaotic orbits. However, as the value of the flattening parameter increases and the galaxy becoming less prolate, the percentage of chaos falls almost linearly and is zeroed when the shape of the elliptical galaxy is spherical $(b = 1)$. The rate of the 2:1 type a resonant orbits increases when $0.1 \leq b \leq 0.5$, while for higher values of the flattening parameter this tendency is reversed. The percentage of box orbits on the other hand, start to grow only at mediocre values of the flattening parameter $(b > 0.4)$ and box orbits become the most populated family when $b > 0.8$, occupying about the two thirds of the entire phase plane. It is evident, that in prolate elliptical galaxies, varying the flattening parameter mainly shuffles the orbital content of the the 2:1 type b,c, 3:1 and 6:3 resonant orbits, whose percentages present fluctuations around 5\%. On the contrary, the 10:5 and all other resonant orbits seem to be immune to changes of the flattening parameter as their rates, at most of the studied cases, remain at very low values less than 5\%. Thus, taking into account all the above, we could say that in prolate elliptical galaxy models the flattening parameter $b$ influences mostly box, 2:1 type a and chaotic orbits. In fact, a large portion of 2:1 type a and chaotic orbits turns into box as the galaxy becomes less prolate, or in other words more spherical.

Of particular interest, would be to determine how the flattening parameter of the elliptical galaxy influences the position of the parent periodic orbits of the different families of orbits shown in the grids of Fig. \ref{gridsP}(a-f). For this purpose, we shall use the theory of periodic orbits \cite{MH92} along with the numerical algorithm developed and applied in \cite{Z13b}. In the left pattern of Figs. \ref{fpo21P} and \ref{fpoP}, we present the evolution of the starting position of the parent periodic orbits of the six basic families of resonant orbits, as a function of the flattening parameter $b$. The evolution of the 2:1 (type a and b), the 6:3 and the 10:5 families is two-dimensional since the corresponding starting position $(R_0,0)$ of all these families lies on the $R$ axis. On the contrary, studying the evolution of the 2:1 type c and 3:1 families of periodic orbits is indeed a real challenge due to the peculiar nature of their starting position $(R_0,\dot{R_0})$. In order to visualize the evolution of these families, we need three-dimensional plots taking into account the simultaneous relocation of the coordinate $R_0$ and the velocity $\dot{R_0}$.

The stability of the periodic orbits can be derived from the elements of the monodromy matrix $X(t)$ as follows:
\begin{equation}
\rm S.I. = {\rm Tr} \left[X(t)\right] - 2,
\end{equation}
where Tr stands for the trace of the matrix, and S.I. is the \emph{stability index}. The periodic orbits are stable if only the stability parameter (S.I.) is between -2 and +2. For each set of values of the flattening parameter $b$, we first located, by means of an iterative process, the position of the parent periodic orbits. Then, using these initial conditions we integrated numerically the variational equations (\ref{vareq}) in order to construct the matrix $X$, from which the stability index was calculated. The evolution of the corresponding stability index (S.I.) of the families of periodic orbits as a function of the flattening parameter $b$ is given in the right pattern of Figs. \ref{fpo21P} and \ref{fpoP}.

It is evident from Figs. \ref{fpo21P} and \ref{fpoP}, that the flattening parameter $b$ plays a key role to the stability of periodic orbits since we observe a strong interplay between stability and instability at many resonant families of orbits, as the value of $b$ varies along the interval $0.1 \leq b \leq 1$. The 2:1 type a resonant family is the dominant family and its stability is completely unaffected by the flattening parameter; 2:1 banana-types orbits remain stable throughout the range of values of $b$. Things however are quite different in the case of the 2:1 type b family where the periodic orbits become unstable at the interval $0.3911 \leq b \leq 0.5862$ and also when $b =  0.8596$. This justifies the absence of the 2:1 type b orbits in the grid shown in Fig. \ref{gridsP}c where $b = 0.5$. Fig. \ref{fpo21P} clearly indicates, however, that the resonance is indeed present, although evidently deeply buried in the chaotic sea due to its unstable nature. On the contrary, the 2:1 type c family contains only stable periodic orbit but exists only when $b \leq 0.5862$. Similar behavior exhibits the 3:1 resonant family which is terminated when $b = 0.6704$. Here we should notice, that both resonant families (2:1 type c and 3:1) are composed entirely of stable periodic orbits and cease to exist abruptly without displaying first any sign of instability. The stability index of the 6:3 resonant family on the other hand, performs significant fluctuations resulting to multiple transitions from stability to instability and vice versa. In particular, when $0.1049 \leq b \leq 0.1157$, $0.2088 \leq b \leq 0.3869$ and $0.7534 \leq b \leq 0.7712$ the 6:3 periodic orbits are unstable, while the 6:3 family ends at $b = 0.7712$. The majority of 10:5 resonant periodic orbits are stable apart from when $0.3912 \leq b \leq 0.5863$ and $0.8504 \leq b \leq 0.9667$ where highly unstable (S.I. $\ll -2$, reaching about -100) periodic orbits appear. The 10:5 resonant family is terminated at $b = 0.9667$.

\subsection{Oblate elliptical galaxy model}
\label{obl}

\begin{table}
\begin{center}
   \caption{ypes and initial conditions of the oblate elliptical galaxy model orbits shown in Figs. \ref{orbsO}(a-g). In all cases, $z_0 = 0$, $\dot{z_0}$ is found from the energy integral, Eq. (\ref{ham}), while $T_{\rm per}$ is the period of the resonant parent periodic orbits.}
   \label{table2}
   \setlength{\tabcolsep}{2.0pt}
   \begin{tabular}{@{}lcccc}
      \hline
      Figure & Type & $R_0$ & $\dot{R_0}$ & $T_{\rm per}$  \\
      \hline
      \ref{orbsO}a & box        &  1.01000000 &  0.00000000 &          - \\
      \ref{orbsO}b & 2:1 banana &  8.43809646 &  0.00000000 &  2.63757246 \\
      \ref{orbsO}c & 1:1 linear &  2.15566054 & 37.26336697 &  1.64882253 \\
      \ref{orbsO}d & 3:2 boxlet &  1.16804695 & 22.21695505 &  4.92122670 \\
      \ref{orbsO}e & 4:3 boxlet & 14.80802308 &  0.00000000 &  6.57229225 \\
      \ref{orbsO}f & 8:5 boxlet & 12.57043007 &  0.00000000 & 13.24880588 \\
      \ref{orbsO}g & chaotic    &  0.27000000 &  0.00000000 &          - \\
      \hline
   \end{tabular}
\end{center}
\end{table}

\begin{figure*}[!tH]
\centering
\resizebox{0.7\hsize}{!}{\includegraphics{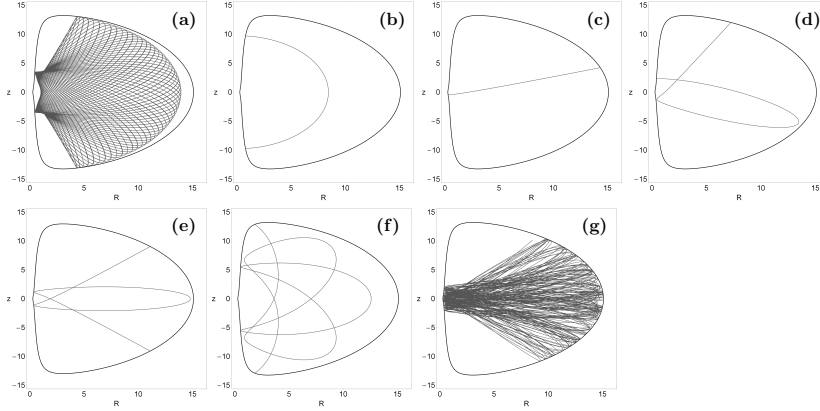}}
\caption{Orbit collection of the seven basic types in the oblate elliptical galaxy model: (a) box orbit; (b) 2:1 banana-type orbit; (c) 1:1 linear orbit; (d) 3:2 boxlet orbit; (e) 4:3 boxlet orbit; (f) 8:5 boxlet orbit; (g) chaotic orbit.}
\label{orbsO}
\end{figure*}

Our numerical investigation shows that in our oblate elliptical galaxy model there are seven main types of orbits: (a) box orbits, (b) 1:1 linear orbits, (c) 2:1 banana-type orbits, (d) 3:2 resonant orbits, (e) 4:3 resonant orbits, (f) 8:5 resonant orbits and (g) chaotic orbits. It is worth noticing, that three of the basic resonant families, that is the 1:1, 3:2 and 4:3 are now present in the oblate elliptical galaxy model. As we emphasised in the previous case, every resonance $n:m$ is expressed in such a way that $m$ is equal to the total number of islands of invariant curves produced by the corresponding orbit in the $(R,\dot{R})$ phase plane. In Fig. \ref{orbsO}(a-g) an example of each of the six basic types of regular orbits, plus an example of a chaotic one are depicted. We observe, that the orbits in the oblate elliptical model obtain considerable lower values of the $z$ coordinate ($z_{max} \simeq 12$ kpc), thus staying close to the galactic plane, while on the other hand, the orbits of the prolate elliptical model shown in Fig. \ref{orbsP}(a-h) obtain high values of $z$ ($z_{max} \simeq 20$ kpc). For all orbits shown in Fig. \ref{orbsO}, we set $b = 1.7$, apart from the orbit shown in Fig. \ref{orbsO}e where $b = 1.9$. The orbits shown in Figs. \ref{orbsO}a and \ref{orbsO}g were computed until $t = 100$ time units, while all the parent periodic orbits were computed until one period has completed. In Table \ref{table2} we provide the type and the exact initial conditions for each of the depicted orbits; for the resonant cases, the initial conditions and the period $T_{\rm per}$ correspond to the parent periodic orbits and are given with precision of eight decimal digits.

\begin{figure*}[!tH]
\centering
\resizebox{0.8\hsize}{!}{\includegraphics{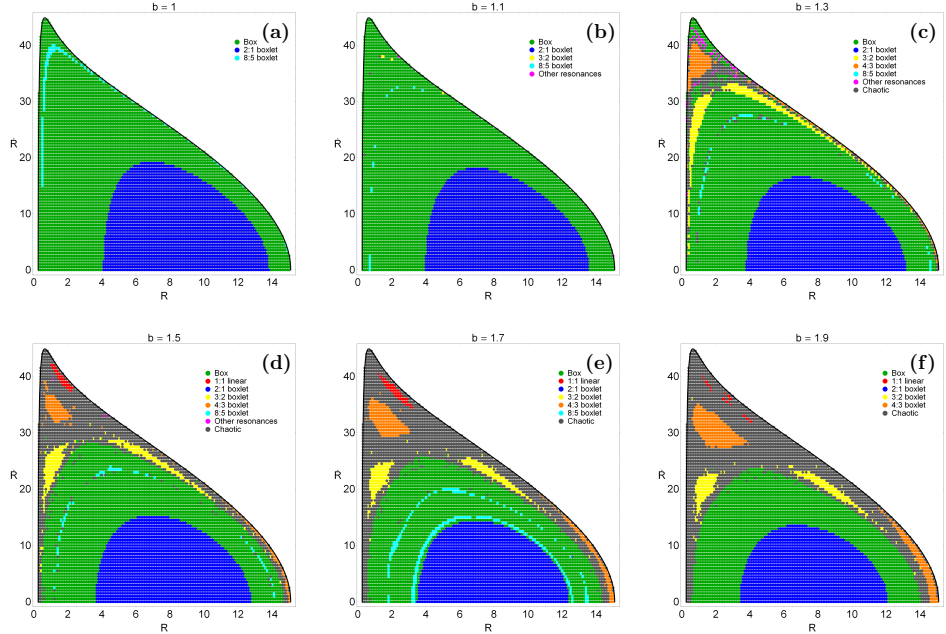}}
\caption{Orbital structure of the $(R,\dot{R})$ phase plane of the oblate elliptical galaxy model for different values of the flattening parameter $b$.}
\label{gridsO}
\end{figure*}

In an attempt to reveal how the flattening parameter $b$ of the elliptical galaxy influences the level of chaos and the overall orbital structure in general, we let it vary while fixing the values of all the other parameters in our oblate elliptical galaxy model. As already said, we fixed the values of all the other parameters and integrate orbits in the meridional plane for the set $b = \{1,1.1,1.3, ..., 1.9\}$. In all cases the energy value was set to $-1050$. Once the values of the parameters were chosen, we defined a set of initial conditions as described in Section \ref{compmeth} and integrated the corresponding orbits computing the SALI of the orbits and then classifying regular orbits into different families.

Six grids of initial conditions $(R_0,\dot{R_0})$ of orbits that we have classified for different values of the flattening parameter $b$ of the oblate elliptical galaxy are shown in Figs. \ref{gridsO}(a-f). By inspecting these grids, we can easily identify all the different regular families by the corresponding sets of islands which are produced in the phase plane. In particular, we see the six main families of orbits already mentioned: (i) 2:1 banana-type orbits correspond to the central periodic point; (ii) box orbits that are situated mainly outside of the 2:1 resonant orbits; (iii) 1:1 open linear orbits form the double set of elongated islands in the outer parts of the phase plane; (iv) 3:2 resonant orbits form the double set of islands above the box orbits; (v) 4:3 resonant orbits correspond to the outer triple set of islands shown in the phase plane; (vi) 8:5 resonant orbits producing the set of five islands which are situated mainly inside the box area. Here we must point out, that those grids show only the $\dot{R} > 0$ part of the phase plane therefore, in many resonant cases not all the islands of the corresponding sets are shown, i.e., one of the two 1:1 islands is shown, two of the three 4:3 islands are present, three of the five 8:5 islands are depicted, etc.

Undoubtedly, the orbital structure of the phase plane in the oblate elliptical galaxy models differs greatly from that of the prolate elliptical models. We observe, in Fig. \ref{gridsO}b where $b = 1.1$, that a minor deviation from spherical symmetry has not significant effect on the orbital structure. Once more, as in the case of the spherical elliptical galaxy $(b = 1)$ the vast majority of the phase plane is covered by box and 2:1 banana-type orbits. Nevertheless, with a much closer look one may distinguish lonely points in the grid corresponding to 3:2 and higher resonant orbits. Things however, change drastically as the value of the flattening parameter increases and the elliptical galaxy is being more oblate. In Fig. \ref{gridsO}c where $b = 1.3$ we see that a weak chaotic layer emerges at the outer parts of the phase plane and the 4:3 resonance along with it. Moreover, the islands of the 3:2 resonance look more well-structured, while several secondary resonances like the 5:4, 6:5 and 7:5 are scattered all over inside the chaotic layer. It is seen, that for larger values of the flattening parameter $b \geq 1.5$ the chaotic layer transforms into a unified chaotic sea and the 1:1 resonance joins the others. When the elliptical galaxy is highly oblate, that is when $b = 1.9$ (see Fig. \ref{gridsO}f) there is no evidence whatsoever of the 8:5 and higher resonances.

\begin{figure}[!tH]
\includegraphics[width=\hsize]{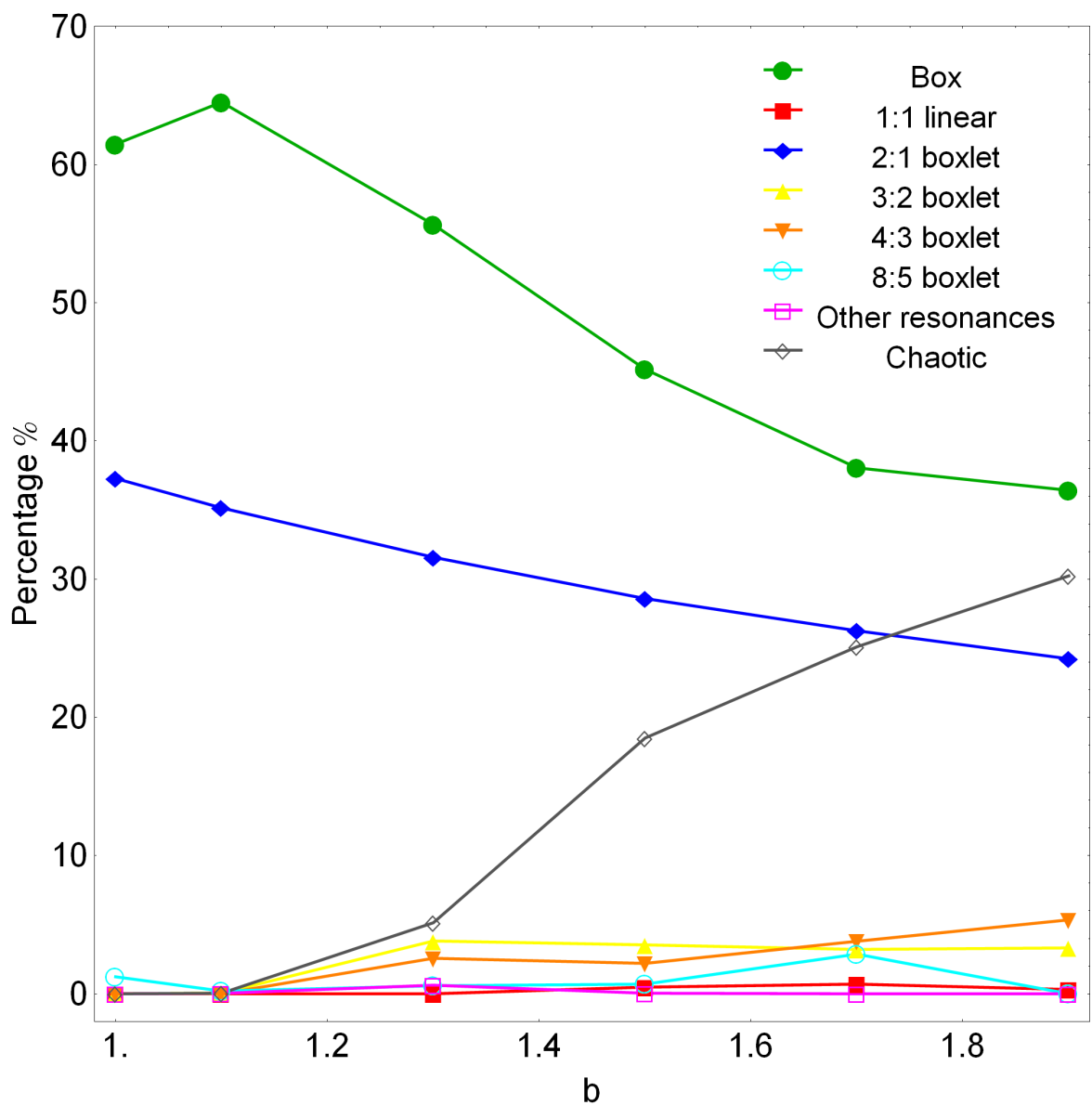}
\caption{Evolution of the percentages of the different types of orbits in our oblate elliptical galaxy model, when varying the flattening parameter $b$.}
\label{percsO}
\end{figure}

\begin{figure}[!tH]
\includegraphics[width=\hsize]{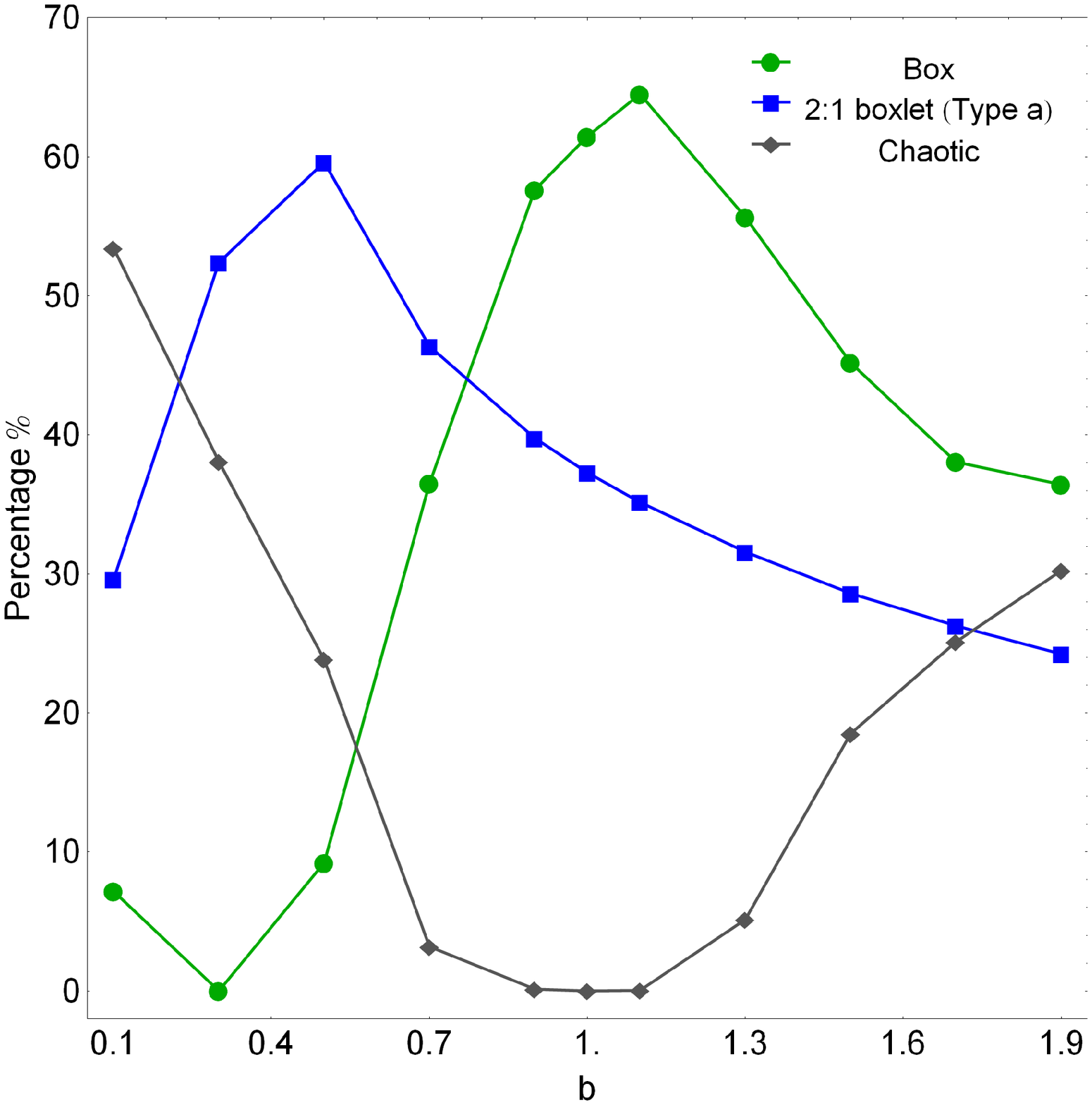}
\caption{Combined evolution of the percentages of the box, 2:1 type a and chaotic orbits in our prolate/oblate elliptical galaxy model, as a function of the flattening parameter $b$.}
\label{percsC}
\end{figure}

In the following Fig. \ref{percsO} we present the resulting percentages of the chaotic orbits and of also of the main families of regular orbits as the value of the flattening parameter $b$ varies. It can be seen, that the motion of stars in oblate elliptical galaxies with small deviation form spherical symmetry $(b < 1.3)$, is almost entirely regular, being the box orbits the all-dominant type. The percentage of box orbits is however reduced as the value of the flattening parameter is increased (the elliptical galaxy becomes more oblate), although they always remain the most populated family. It is also seen, that the percentage of the 2:1 banana-type resonant orbits exhibits a minor and almost linear decrease with increasing $b$. The chaotic orbits on the other hand, start to grow rapidly as soon as the galaxy is oblate and sufficiently away from axially symmetry $(b > 1.1)$. At the value of the flattening parameter studied $(b = 1.9)$ the percentages of box and chaotic orbits seem to tend to a common value (around 35\%), thus sharing seven tenths of the entire phase plane. Furthermore, the percentages of the 3:2 and 4:3 resonant orbits are slightly amplified with increasing $b$. On the contrary, all the other resonant families of orbits (1:1, 8:5 and other types) are immune to changes of the flattening parameter, since their rates remain almost unperturbed and at extremely low values (less than 5\% throughout). Therefore, from Fig. \ref{percsO} one may conclude, that flattening parameter in oblate elliptical galaxies mostly affects box, 2:1 banana-type and chaotic orbits. In fact, a portion of box orbits turns into chaotic as the galaxy becomes more oblate.

We seen, that box, 2:1 type a and of course chaotic orbits are common in both prolate and oblate elliptical galaxy models. In fact, these are the main types of orbits which are the building blocks of the orbital structure of the galaxy. Therefore, we decided to combine the corresponding results presented in subsections \ref{prl} and \ref{obl} thus showing in Fig. \ref{percsC} the complete evolution of the percentages of these orbits as a function of the flattening parameter $b$. We observe, that each type of orbits prevails in different range of the flattening parameter. In particular, box orbits is the dominant type when $b > 0.8$, 2:1 type a orbits prevail when $0.2 < b < 0.7$, while chaotic orbits take over only when $b < 0.2$. Thus, we may argue that in general terms, in flattened (prolate and oblate) elliptical galaxy models the majority of orbits is regular.

\begin{figure*}[!tH]
\centering
\resizebox{0.5\hsize}{!}{\includegraphics{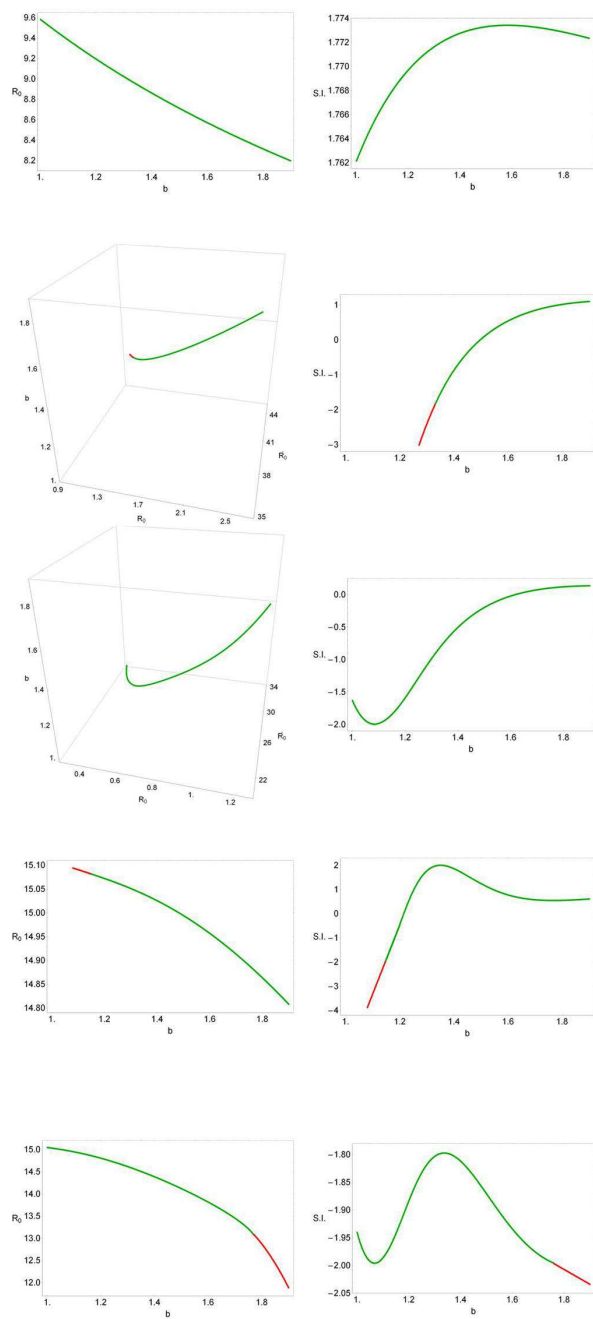}}
\caption{(left pattern): Evolution of the starting position $(R_0,\dot{R_0})$ of the periodic orbits as a function of the flattening parameter $b$. (right pattern): Evolution of the corresponding stability index (S.I.) of the families of periodic orbits as a function of the flattening parameter $b$. From top to bottom: 2:1 resonant family, 1:1 resonant family, 3:2 resonant family, 4:3 resonant family and 8:5 resonant family. Green dots indicate stable periodic orbits, while red dots correspond to unstable periodic orbits.}
\label{fpoO}
\end{figure*}

We continue our exploration, by presenting how the variation in the flattening parameter of the oblate elliptical galaxy influences the position of the different families of periodic orbits shown in the grids of Fig. \ref{gridsO}(a-f). To make things simple, we follow the same method used previously in the case of the prolate elliptical galaxy. In the left pattern of Fig. \ref{fpoO} the evolution of the starting position of the parent periodic orbits of the five basic families of resonant orbits is given, while in the right pattern we present the evolution of the corresponding stability index (S.I.) of the families of periodic orbits as a function of the flattening parameter $b$. We observe, that the evolution of the 2:1, 4:3 and 8:5 families shown in Fig. \ref{fpoO}, is two-dimensional since the starting position $(R_0,0)$ of these families lies on the $R$ axis. On the contrary, for the evolution of the 1:1 and 3:2 families, we need three-dimensional plots thus following simultaneous the relocation of $R_0$ and $\dot{R_0}$ as the flattening parameter $b$ varies.

It can be seen in Fig. \ref{fpoO}, that as the different resonant families evolve some stable periodic orbits (green dots) become unstable (red dots) and vice versa. One of the most stable resonant families is the 2:1 family which remains stable throughout the range of the values of the flattening parameter. Similarly, the 3:2 resonant family remains also stable throughout, even though the value of the corresponding stability index approaches very close to the low limit of stability (S.I. = -2) when $b = 1.0793$. All the other resonant families, on the other hand, contain a mixture of stable and unstable periodic orbits. The 1:1 resonant family remains stable only when $1.3221 \leq b \leq 1.9$, while for $b < 1.3221$ all the periodic orbits are unstable. Moreover, the 1:1 family ends at $b = 1.2714$, thus without reaching the low limit of the flattening parameter $(b = 1)$. It is worth noticing that, even though an unstable 1:1 periodic orbit exists when $b = 1.3$, there is no evidence of the 1:1 resonance in the corresponding grid of Fig. \ref{gridsO}c. In the case of the 4:3 resonant family, the periodic orbits are stable when $b \geq 1.1461$, while for lower values of the flattening parameter they become unstable and the entire family ceases to exist at $b = 1.0798$. As in the case of the 1:1 resonance, in the grid of Fig. \ref{gridsO}c the unstable 4:3 periodic orbit is absent, evidently deeply buried in the vast area occupied by box orbits. Finally we observe, that the 8:5 resonant family covers the entire range of the values of $b$, being stable however, only when $b \leq 1.7731$. At the higher value of the flattening parameter $(b = 1.9)$ the 8:5 periodic orbit is unstable and consequently, undetected in the corresponding grid shown in Fig. \ref{gridsO}f. Therefore it becomes clear, that the flattening parameter in oblate elliptical galaxy models influences significantly not only the existence but also the stability of the different regular families.

Before closing this Section, we would like to stress out that usually the 1:1 resonance is the hallmark of the loop orbits, at which both coordinates oscillating with the same frequency in their main motion and their mother orbit is a closed loop orbit. Furthermore, when the oscillations are in phase, the 1:1 resonant orbit degenerates into a linear orbit (the same as in Lissajous figures made with two oscillators). In our meridional plane, however, the 1:1 orbits do not have the shape of a loop. In fact, their parent orbit is linear (as in Fig. \ref{orbsO}c) and thus they do not have a hollow (referring to the meridional plane) but fill a region around the linear parent orbit, always oscillating along the $R$ and $z$ directions with the same frequency. We decided to call these orbits ``1:1 linear open orbits" in order to differentiate them from true meridional plane loop orbits, which do have a hollow and also rotate always in the same sense.

\section{Discussion and conclusions}
\label{disc}

In the present work, we used a new analytic, axisymmetric galactic gravitational model which embraces the general features of a flattened elliptical galaxy containing a dense, massive nucleus with an additional spherical dark matter halo component. The choice of the model potential for the description of the elliptical galaxy was made mainly taking into account that the vast majority of the observed galaxies are flattened; either prolate or oblate. In order to simplify our study, we chose to work in the meridional plane $(R,z)$. Our aim was to investigate how influential is the flattening parameter on the level of chaos and also on the distribution of regular families in our elliptical galaxy model. Our results strongly suggest, that the level of chaos and the distribution of regular families is indeed very dependent on the particular shape of the galaxy. We kept the values of all the other parameters constant because our main objective was to investigate the influence of the flattening parameter $b$ on the percentages of the orbits.

It is beyond any doubt, that knowing the overall orbital structure in a galactic model is an issue of significant importance. However, the vast majority of the existed literature on this aspect is limited distinguishing mostly between regular and chaotic orbits, while only a handful of them dive deeper separating regular orbits into different families. Therefore, we decided to conduct in a series of papers \cite{ZC13,CZ13,ZCar13,ZCar14} a thorough orbit classification in several types of galactic potentials, in an attempt to shed some light on how the basic involved parameters of each system influence the percentages not only of the chaotic but also of the different types of regular orbits. We believe, that the presented outcomes can provide interesting information regarding the structure and properties of flattened elliptical galaxies.

An elliptical galaxy with a spherical nucleus and an additional dark matter halo component is surely a very complex entity and, therefore, we need to assume some necessary simplifications and assumptions in order to be able to study mathematically the orbital behavior of such a complicated stellar system. For this purpose, our model is intentionally simple and contrived, as all the other models of this series, in order to give us the ability to study all the different aspects regarding the kinematics and dynamics of the model. Nevertheless, contrived models can provide an insight into more realistic stellar systems, which unfortunately are very difficult to be studied, if we take into account all the astrophysical aspects (i.e. gas, spirals, mergers, etc). Self-consistent models on the other hand, are mainly used when conducting $N$-body simulations. However, this application is entirely out of the scope of the present research. Once again, we have to point out that the simplicity of our model is necessary, otherwise it would be extremely difficult, or even impossible, to apply the extensive and detailed numerical calculations presented in this study. Similar gravitational models with the same limitations and assumptions were used successfully several times in the past in order to investigate the orbital structure in much more complicated galactic systems \cite[e.g.,][]{CI09,C12,Z12b,Z13a}.

Since a distribution function of the galaxy model was not available so as to use it for extracting the different samples of orbits, we had to follow an alternative path. So, for determining the regular or chaotic nature of motion in our models, we chose, for each set of values of the flattening parameter, a dense grid of initial conditions in the $(R,\dot{R})$ phase plane, regularly distributed in the area allowed by the value of the orbital energy $E$. Each orbit was integrated numerically for a time period of $10^4$ time units (10 billion yr), which corresponds to a time span of the order of hundreds of orbital periods but of the order of one Hubble time. The particular choice of the total integration time was made in order to eliminate sticky orbits (classifying them correctly as chaotic orbits) with a stickiness at least of the order of one Hubble time. Then, we made a step further, in an attempt to distribute all regular orbits into different families. Therefore, once an orbit has been characterized as regular applying the SALI method, we then further classified it using a frequency analysis method. This method calculates the Fourier transform of the coordinates and velocities of an orbit, identifies its peaks, extracts the corresponding frequencies and then searches for the fundamental frequencies and their possible resonances.

Our numerical investigation points out, that the flattening parameter of the elliptical galaxy has a key role and affects significantly the orbital content of the system. The main results of our research can be summarized as follows:
\begin{enumerate}
 \item In our galaxy models several types of regular orbits exist, while there is also an extended chaotic domain separating the areas of regularity. In particular, in the case of prolate models three different types of 2:1 resonant orbits coexist, while two bifurcated families, that is the 6:3 and 10:5 orbits, are also appear. In oblate galaxy models on the other hand, a large variety of resonant orbits (i.e. 1:1, 2:1, 3:2, 4:3, 8:5 and higher resonant orbits) are present thus making the orbital structure more rich. Here we must clarify, that by the term ``higher resonant orbits" we refer to resonant orbits with a rational quotient of frequencies made from integers $> 5$, which of course do not belong to the main families.
 \item There is a strong correlation between the percentages of most types of orbits and the value of the flattening parameter $b$. In the case of a prolate elliptical galaxy $0.1 \leq b < 0.9$ we found that the flattening parameter affects mostly the percentages of box, 2:1 type a and chaotic orbits. It was observed, that as the galaxy becomes less prolate the rate of box orbits grows at the expense of 2:1 type a and chaotic orbits. In oblate $(1.1 \leq b < 1.9)$ elliptical galaxies once more, the flattening parameter influences mainly box, 2:1 banana-type and chaotic orbits. However this time, a portion of box orbits turns into chaotic as the galaxy becomes more oblate. When the elliptical galaxy is spherical $(b = 1)$ the motion is entirely regular as box and 2:1 type a orbits swarm the phase plane.
 \item Our investigation revealed that in flattened elliptical galaxy models, there are stable as well as unstable periodic orbits. We conducted a thorough numerical analysis on how the flattening parameter influences the position of the parent periodic orbits of each regular family. We found, that the periodic orbits of the 2:1 type a, 2:1 type c, 3:1 and 3:2 families remain stable throughout the entire range of the values of $b$. On the contrary, the 1:1, 2:1 type b, 4:3, 6:3, 8:5 and 10:5 families of periodic orbits contain not only stable but also a considerable amount of unstable periodic orbits. Therefore, we concluded that the flattening parameter $b$ affects substantially the stability of the regular families of orbits, hinting at a deep interplay between chaos and proportion of regular families.
\end{enumerate}

The regular or chaotic nature of orbits in axially symmetric elliptical galaxies has been studied in numerous previous works. \cite{KC01} used a simple logarithmic potential in order to model the properties of motion of stars moving in the meridional plane of an elliptical galaxy with a dense nucleus. It was found, that the flattening parameter of the galaxy as well as the value of the angular momentum play a fundamental role to the transition from regularity to chaos. In particular, it was observed that low angular momentum stars are in chaotic orbits in highly flattened elliptical galaxies. A similar logarithmic potential with an additional perturbing term was utilized in \cite{CP03} where interesting correlations between the perturbing term nd other physical quantities of the dynamical system were presented. \cite{CZ11} expanded the work of \cite{KC01} by adding a spherical nucleus in the total potential and investigated how the mass of the nucleus and the angular momentum influence the character of orbits in a prolate elliptical galaxy. Some additional relations connecting other parameters with the level of chaos were also found.

A composite logarithmic potential describing the motion of stars in elliptical as well as in disk galaxies, by suitably choosing the dynamical parameters was introduced in \cite{Z11}. In the case of the elliptical galaxy, it was observed that the flattening parameter affects not only the percentage but also the degree of chaoticity. The numerically obtained results were supported and confirmed by semi-theoretical arguments. This modified logarithmic potential was deployed in \cite{ZCar13} in an attempt to explore the effects of the fraction of dark matter on the orbital structure in elliptical galaxies. We found, that in elliptical galaxy models the motion of stars is highly chaotic depopulating the main regular families (i.e. the box and the 2:1 resonant orbits). In a recent work, \cite{CZ13} showed that in axisymmetric elliptical galaxies mainly box and chaotic orbits are influenced by the portion of dark matter. In fact, box orbits are the all dominant family when the amount of dark matter is low, however, the rate of chaotic orbits quickly grows as the amount of dark matter is increased, thus suppressing the percentage of box orbits, although they remain the most populated family.

We consider the results of the present research as an initial effort and also a promising step in the task of understanding the orbital structure of flattened elliptical galaxies. Taking into account that our outcomes are encouraging, it is in our future plans to modify properly our galaxy model in order to expand our investigation into three dimensions, thus exploring how the flattening parameter influences the nature of the three-dimensional (3D) orbits in triaxial elliptical galaxies. Furthermore, of particular interest would be to reveal the entire network of periodic orbits, thus shedding some light to the evolution of the periodic orbits as well as their stability when varying all the available parameters of the dynamical system.

\section*{Acknowledgments}

E.E.Z. would like to thank Dr. D.D. Carpintero for his valuable and continuous contribution to our efforts to further refinement and improvement of the orbit classification code.

\label{pagefin}
\end{document}